\documentclass[preprint,showpacs,onecolumn,nofootinbib]{revtex4}
\pdfoutput=1
\usepackage[colorlinks=true,linkcolor=blue,urlcolor=blue,filecolor=black,citecolor=red,pdfstartview=FitV,pdftitle={},pdfsubject={},pdfkeywords={},pdfpagemode=None,bookmarksopen=true]{hyperref}
\usepackage{graphicx}
\usepackage{amsfonts}
\usepackage{amssymb,ulem}
\usepackage{color}%
\usepackage{dcolumn}
\usepackage{amsmath}
\usepackage{epsfig}
\usepackage{graphics}
\usepackage[active]{srcltx}
\usepackage{epstopdf}
\usepackage{shuffle}
\usepackage[utf8]{inputenc}

\newcommand{\bea}{\begin{eqnarray}}
\newcommand{\eea}{\end{eqnarray}}
\newcommand{\bean}{\begin{eqnarray*}}
\newcommand{\eean}{\end{eqnarray*}}
\newcommand{\nn}{\nonumber \\}

\def\W #1{\widetilde{#1}}

\def\Tr{\mathop{\rm Tr}}

\def\eref#1{(\ref{#1})}

\def\a{{\alpha}}

\def\b{{\beta}}

\def\Label#1{\label{#1}%
  \smash{\hbox to0pt{\raise1ex\hbox{\tiny[#1]}\hss}}}

\begin{document}

\baselineskip=0.6 cm
\title{Expanding single trace YMS amplitudes with gauge invariant coefficients}
\author{Fang-Stars Wei}
\email{mx120220339@stu.yzu.edu.cn}
\author{Kang Zhou}
\email{zhoukang@yzu.edu.cn}
\affiliation{Center for Gravitation and Cosmology, College of Physical Science and Technology, Yangzhou University, Yangzhou, 225009, China}

\begin{abstract}
\baselineskip=0.6 cm

In this note, we use the new bottom up method based on soft theorems to construct the expansion of single-trace Yang-Mills-scalar amplitudes recursively. The resulted expansion manifests the gauge invariance for any polarization carried by external gluons, as well as the permutation symmetry among external gluons. Our result is equivalent to that found by Clifford Cheung and James Mangan via the so called covariant color-kinematic duality approach.

\end{abstract}

\maketitle


\section{Introduction}
\label{sec-intro}

The investigations of $S$-matrix in the past decade revealed deep connections among amplitudes of various different theories, which are invisible upon inspecting traditional Feynman rules.
For instance, tree level gravitational (GR) and Yang-Mills (YM) amplitudes are related by the so called Kawai-Lewellen-Tye (KLT) relation \cite{Kawai:1985xq}, and Bern-Carrasco-Johansson (BCJ) color-kinematic duality \cite{Bern:2008qj,Chiodaroli:2014xia,Johansson:2015oia,Johansson:2019dnu}. In the well known CHY formalism \cite{Cachazo:2013gna,Cachazo:2013hca, Cachazo:2013iea, Cachazo:2014nsa,Cachazo:2014xea}, tree amplitudes for a large variety of theories can be generated from tree GR amplitudes through the compactifying, squeezing and the generalized dimensional reduction procedures. Similar unifying relations were proposed by introducing appropriate differential operators, which transmute tree GR amplitudes to others \cite{Cheung:2017ems,Zhou:2018wvn,Bollmann:2018edb}. At the same time, another type of relations also caused attentions, which can be called the expansions of amplitudes, namely, tree amplitudes of one theory can be expanded to tree amplitudes of other theories \cite{Fu:2017uzt,Teng:2017tbo,Du:2017kpo,Du:2017gnh,Feng:2019tvb,Zhou:2019gtk,Zhou:2019mbe}. Indeed, the expansions serve as the dual picture of differential operators, as interpreted in \cite{Zhou:2019mbe}. In the dual web, tree amplitudes for a wide range of theories can be expanded to tree bi-adjoint scalar (BAS) amplitudes.

In this note, we focus on the expansion of the tree single-trace Yang-Mills-scalar (YMS) amplitudes \cite{Stieberger:2016lng,Schlotterer:2016cxa,Chiodaroli:2014xia,Chiodaroli:2017ngp,DelDuca:1999rs,Nandan:2016pya,delaCruz:2016gnm}, due to the special role of these amplitudes. Both tree YM and BAS amplitudes can be regarded as special cases of tree single-trace YMS amplitudes. When performing the squeezing procedure in CHY framework, or acting differential operators, the YM amplitudes are transmuted to single-trace YMS amplitudes, then end with pure BAS ones. These connections can be extended to tree GR amplitudes, tree YM amplitudes, and single-trace tree EYM amplitudes directly \cite{Cachazo:2013iea,Cachazo:2014nsa,Cachazo:2014xea,Zhou:2018wvn,Bollmann:2018edb}.
Furthermore, single-trace tree YMS amplitudes exhibit double copy structure and color-kinematic duality in an elegant manner \cite{Cachazo:2013iea,Cachazo:2014nsa,Cachazo:2014xea,Cheung:2021zvb,Chiodaroli:2014xia,Chiodaroli:2017ngp}.
The single-trace YMS amplitudes which describe the scattering of massless gluons and scalars, can be expanded to YMS ones with less external gluons and more external scalars \cite{Fu:2017uzt,Teng:2017tbo,Du:2017kpo,Du:2017gnh,Feng:2019tvb,Zhou:2022orv}.
Such recursive expansion can be applied iteratively, end with the expansion to pure BAS amplitudes.

In the recursive expansion given in \cite{Fu:2017uzt,Teng:2017tbo,Feng:2019tvb}, a fiducial external gluon is required. This special gluon breaks the manifest permutation symmetry among external gluons: if one re-chose the fiducial gluon, we expect that the new expansion is equivalent to the old one, but such equivalence is hard to prove. Another disadvantage of the expansion in \cite{Fu:2017uzt,Teng:2017tbo,Feng:2019tvb} is related to the gauge invariance. In the expansion in \cite{Fu:2017uzt,Teng:2017tbo,Feng:2019tvb}, the gauge invariance for the polarization vector carried by the fiducial gluon is obscure, namely, if one replace the polarization by the corresponding momentum, the vanishing of amplitude is not manifest. suppose one use this expansion iteratively to expand the YMS amplitude to BAS ones, in the resulting expansion, non of polarizations has the manifest gauge invariance, since each external gluon will play the role of the fiducial gluon once in the iterative process. As well known, the gauge invariance plays the crucial role in the modern S-matrix program, and always hints new understanding and new mathematical structure for scattering amplitudes.
For instance, tree amplitudes in general relativity and Yang-Mills theory, turned out to be completely determined by gauge invariance and singularity structures \cite{Arkani-Hamed:2016rak,Rodina:2016jyz}. Another well known example is that the Britto-Cachazo-Feng-Witten (BCFW) on-shell recursion relation expresses YM amplitudes in gauge invariant formulas which include spurious poles \cite{Britto:2004ap,Britto:2005fq}. These new formulas motivated elegant new constructions of amplitudes, such as Grassmannia representation and Amplituhedron \cite{Arkani-Hamed:2012zlh,Arkani-Hamed:2013jha,Arkani-Hamed:2013kca}. These experiences force us to seek the new expansion with the explicit gauge invariance. After expanding to BAS amplitudes, the BAS basis contributes only poles, thus the gauge invariance for each polarization is completely determined by coefficients. Thus, it is natural to expect a formula of coefficients which manifests the gauge invariance. The above consideration leads to an important question, how to achieve the new expansion which have manifest permutation symmetry among external gluons and the gauge invariance for each polarization of external gluons?

Such new expansion was first found by Clifford Cheung and James Mangan in \cite{Cheung:2021zvb}, via the so called covariant color-kinematic duality method. The approach in \cite{Cheung:2021zvb} is based on the traditional Lagrangian and equations of motion. The very premise of the modern S-matrix program is to bootstrap scattering dynamics without the aid of an action or equation of motion (see for reviews in \cite{Elvang:2013cua,Cheung:2017pzi}). Thus, it is natural to ask whether the new expansion in \cite{Cheung:2021zvb} can be obtained through the bottom up construction which uses only on-shell information? This question is the main motivation for the current short note.

In this note, we reconstruct the expansion in \cite{Cheung:2021zvb}, from a totally different perspective based on universal soft behaviors of massless particles. We first bootstrap the lowest $3$-point tree single-trace YMS amplitudes with only one external gluon, by imposing general principles such as the appropriate mass dimension and Lorentz invariance. Then, we invert the soft theorem for external scalars to construct the expanded single-trace YMS amplitudes with more external scalars, while keeping the number of external gluons to be un-altered. After such construction, we use the well known Bern-Carrasco-Johansson (BCJ) relation \cite{Bern:2008qj,Chiodaroli:2014xia,Johansson:2015oia,Johansson:2019dnu} to turn the resulted expansion to the new formula which manifests the gauge invariance. Next, we invert the sub-leading soft theorem for external gluons to generate the expanded single-trace YMS amplitudes with more external gluons. Such procedure inserts external gluons to the original amplitude in a manifestly gauge invariant pattern, thus
the explicit gauge invariance will be kept if the starting point is expressed in a gauge invariant form. The manifest permutation symmetry among external gluons are also kept at each step.

The remainder of this note is organized as follows. In section.\ref{sec-background}, we rapidly introduce the necessary background including the expansion of tree massless amplitudes to BAS amplitudes, the recursive expansion of single-trace YMS amplitudes, as well as soft theorems for external scalars and gluons. In section.\ref{sec-expan-YMS}, we construct the expansion in \cite{Cheung:2021zvb} by using our recursive method based on inverting soft theorems. Then, we end with a brief summery in section.\ref{sec-conclusion}.

\section{Background}
\label{sec-background}

For readers' convenience, in this section we give a brief review of necessary background. In subsection.\ref{subsecexpand}, we introduce the tree level amplitudes of bi-adjoint scalar (BAS) theory, as well as expansions of tree amplitudes to BAS amplitudes. In subsection.\ref{subsec-expan-YMS}, we give the recursive expansion of Yang-Mills-scalar (YMS) amplitudes. In subsection.\ref{subsec-soft}, we review the soft theorems for external scalars and gluons.

\subsection{Expanding tree level amplitudes to BAS basis}
\label{subsecexpand}

The bi-adjoint scalar (BAS) theory describes the massless bi-adjoint scalar fields $\phi^{Aa}$ with cubic interaction, the Lagrangian is given as
\bea
{\cal L}_{\rm BAS}={1\over2}\,\partial_\mu\phi^{Aa}\,\partial^{\mu}\phi^{Aa}+{\lambda\over3!}\,F^{ABC}f^{abc}\,
\phi^{Aa}\phi^{Bb}\phi^{Cc}\,,
\eea
where the structure constant $F^{ABC}$ and generator $T^A$ satisfy
\bea
[T^A,T^B]=iF^{ABC}T^C\,,
\eea
and the dual algebra encoded by $f^{abc}$ and $T^{a}$ is analogous.

\begin{figure}
  \centering
  \includegraphics[width=6cm]{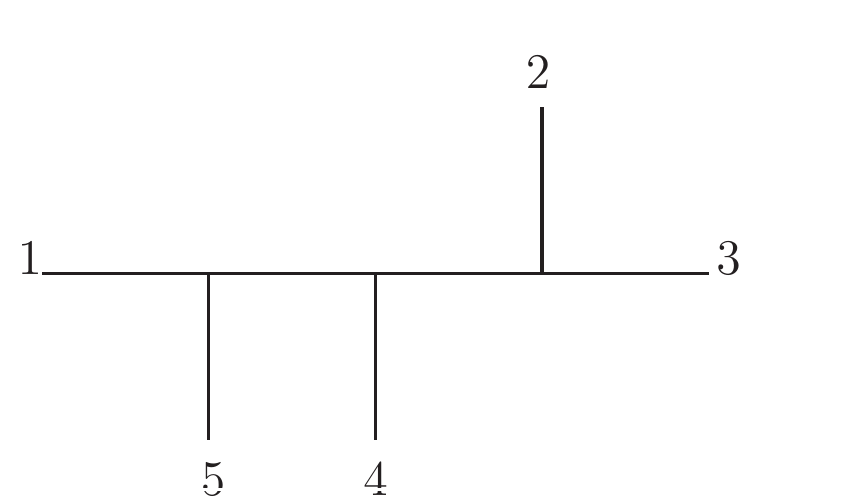}
   \includegraphics[width=6cm]{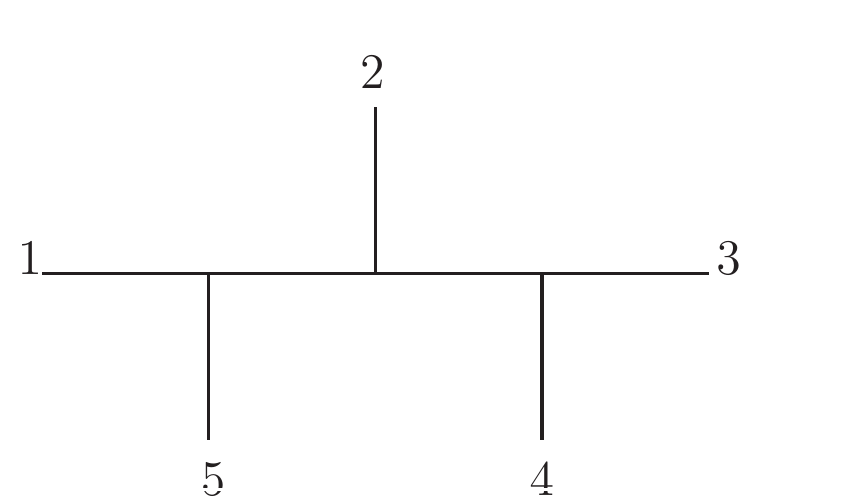}  \\
  \caption{Two $5$-point diagrams}\label{5p}
\end{figure}

Tree amplitudes of this theory only contain propagators for massless scalars. Decomposing the group factors via the standard procedure gives
\bea
{\cal A}_n=\sum_{\sigma\in{\cal S}_n}\,\sum_{\W\sigma_n\in\W{\cal S}_n}\,\Tr[T^{A_{\sigma_1}},\cdots T^{A_{\sigma_n}}]\,\Tr[T^{a_{\W\sigma_1}}\cdots T^{a_{\W\sigma_n}}]\,{\cal A}_{S}(\sigma_1,\cdots,\sigma_n|\W\sigma_1,\cdots,\W\sigma_n)\,,
\eea
where ${\cal A}_n$ denotes the kinematic part of $n$-point amplitude in which the coupling constants are dropped, ${\cal S}_n$ and $\W{\cal S}_n$ are non-cyclic permutations of external legs. In this note, partial amplitudes ${\cal A}_{S}(\sigma_1,\cdots,\sigma_n|\W\sigma_1,\cdots,\W\sigma_n)$ play the crucial role when considering expansions of amplitudes.
Each partial amplitude is simultaneously planar with respect to two color orderings $(\sigma_1,\cdots,\sigma_n)$ and $(\W\sigma_1,\cdots,\W\sigma_n)$. In other words, it is double color ordered. Let us take the $5$-point amplitude ${\cal A}_{\rm S}(1,2,3,4,5|1,4,2,3,5)$ as the example.
In Figure.\ref{5p}, the first diagram is allowed by both two orderings $(1,2,3,4,5)$ and $(1,4,2,3,5)$, while the second one violates the ordering
$(1,4,2,3,5)$. Other diagrams which satisfy the ordering $(1,2,3,4,5)$ are also forbidden by the ordering $(1,4,2,3,5)$, thus the first diagram in Figure.\ref{5p} is the only candidate for the amplitude
${\cal A}_{\rm S}(1,2,3,4,5|1,4,2,3,5)$. Then the amplitude ${\cal A}_{\rm S}(1,2,3,4,5|1,4,2,3,5)$ can be evaluated as
\bea
{\cal A}_{\rm S}(1,2,3,4,5|1,4,2,3,5)={1\over s_{23}}{1\over s_{51}}\,,
\eea
up to an overall sign. The Mandelstam variable $s_{i\cdots j}$ is defined as
\bea
s_{i\cdots j}\equiv k_{i\cdots j}^2\,,~~~~k_{i\cdots j}\equiv\sum_{a=i}^j\,k_a\,,~~~~\label{mandelstam}
\eea
where $k_a$ is the momentum carried by the external leg $a$.

Each double color ordered partial BAS amplitude carries an overall $\pm$ sign, arises from swapping two lines at the common vertex, due to the anti-symmetry of structure constants $F^{ABC}$ and $f^{abc}$. In this note, we choose the convention that the overall sign is $+$ if two orderings carried by the BAS amplitude are the same. For instance, the amplitude ${\cal A}_{\rm S}(1,2,3,4|1,2,3,4)$ carries the overall sign $+$ under the above convention. Notice that this convention is different from that in \cite{Cachazo:2013iea}. The overall signs for other BAS amplitudes with general orderings can be determined by counting the number of flipping \cite{Cachazo:2013iea}.

The double color ordered partial amplitudes can be systematically evaluated by applying the diagrammatical rules proposed by Cachazo, He and Yuan in \cite{Cachazo:2013iea}. We do not introduce this method in the current note, the reader can see this interesting and useful approach in \cite{Cachazo:2013iea}.

\begin{figure}
  \centering
   \includegraphics[width=8cm]{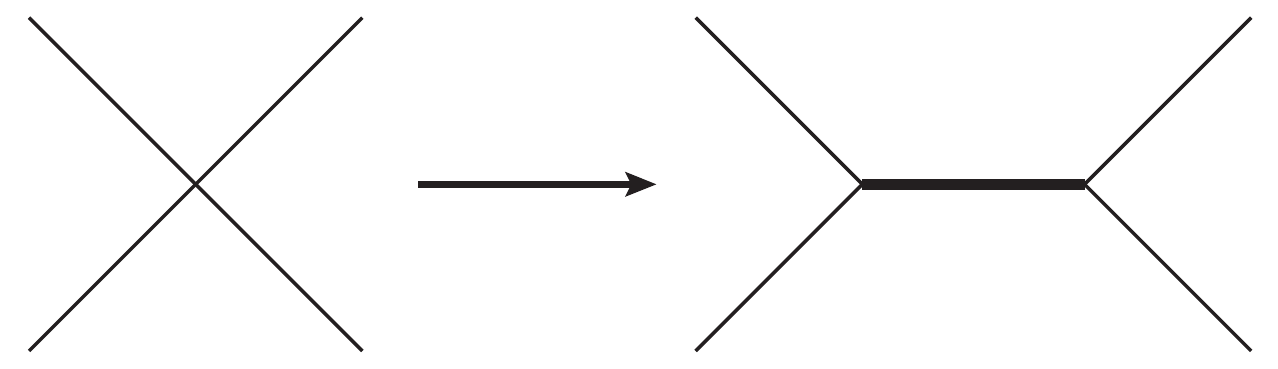}\\
  \caption{Turn the $4$-point vertex to $3$-point ones. The bold line corresponds to the inserted propagator $1/D$. This manipulation turns the original numerator $N$ to $DN$.}\label{3pto4p}
\end{figure}

Tree level amplitudes for massless particles and cubic interactions can be expanded to double color ordered partial BAS amplitudes,
since each associated Feynman diagram can be included in at least one partial BAS amplitude. For higher-point vertices, one can transmute them to cubic ones by inserting the propagator $1/D$ and the numerator $D$ simultaneously, as shown in Figure.\ref{3pto4p}. This manipulation expands any tree amplitude to tree Feynman diagrams with only cubic vertices. Since each Feynman diagram contributes propagators which can be provided by partial BAS amplitudes, accompanied with a numerator, one can conclude that each tree amplitude for massless particles can be expanded to double color ordered partial BAS amplitudes. In such expansions, coefficients are polynomials depend on Lorentz invariants created by external kinematical variables, without any pole.

The expansion requires appropriate basis.
Such basis can be obtained by employing the well known Kleiss-Kuijf (KK) relation \cite{Kleiss:1988ne}
\bea
{\cal A}_{\rm S}(1,\vec{\pmb{\a}},n,\vec{\pmb{\b}}|\sigma_n)=(-)^{|\pmb{\b}|}\,{\cal A}_{\rm S}(1,\vec{\pmb{\a}}\shuffle\vec{\pmb{\b}}^T,n|\sigma_n)\,,~~~\label{KK}
\eea
where $\vec{\pmb{\a}}$ and $\vec{\pmb{\b}}$ are two ordered subsets of external scalars, and $|\pmb{\b}|$ stands for the number of elements included in $\vec{\pmb{\b}}$. The ordered set $\vec{\pmb{\b}}^T$ is the inversion of $\vec{\pmb{\b}}$. For example, $\vec{\pmb{\b}}^T=\{3,2,1\}$ if $\vec{\pmb{\b}}=\{1,2,3\}$. The $n$-point BAS amplitude ${\cal A}_{\rm S}(1,\vec{\pmb{\a}},n,\vec{\pmb{\b}}|\sigma_n)$ at the l.h.s of \eref{KK} carries two orderings, which are encoded by $(1,\vec{\pmb{\a}},n,\vec{\pmb{\b}})$ and $\sigma_n$ respectively. The symbol $\shuffle$ means summing over all
possible shuffles of two ordered sets $\vec{\pmb{\gamma}}_1$ and $\vec{\pmb{\gamma}}_2$, i.e., all permutations in the set $\vec{\pmb{\gamma}}_1\cup \vec{\pmb{\gamma}}_2$ while preserving the orderings
of $\vec{\pmb{\gamma}}_1$ and $\vec{\pmb{\gamma}}_2$. For instance, suppose $\vec{\pmb{\gamma}}_1=\{1,2\}$ and $\vec{\pmb{\gamma}}_2=\{3,4\}$, then
\bea
{\cal A}(\vec{\pmb{\gamma}}_1\shuffle \vec{\pmb{\gamma}}_2)&=&{\cal A}(1,2,3,4)+{\cal A}(1,3,2,4)+{\cal A}(1,3,4,2)\nn
& &+{\cal A}(3,1,2,4)+{\cal A}(3,1,4,2)+{\cal A}(3,4,1,2)\,.~~~~\label{shuffle}
\eea
The analogous KK relation holds for the ordering $\sigma_n$.
The KK relation among partial BAS amplitudes implies that the basis can be chosen as amplitudes
${\cal A}_{\rm S}(1,\sigma_1,n|1,\sigma_2,n)$, with $1$ and $n$ are fixed at two ends in each ordering. Such basis is called the KK BAS basis. Consequently, any tree amplitude for massless particles can be expanded to this basis.
The basis provides poles, while the coefficients contribute numerators.

The double color ordered partial BAS amplitudes also satisfy well known Bern-Carrasco-Johansson (BCJ) relations \cite{Bern:2008qj,Chiodaroli:2014xia,Johansson:2015oia,Johansson:2019dnu}. Here we give the explicit formula of the fundamental BCJ relation,
\bea
(k_s\cdot Y_s)\,{\cal A}_{\rm S}(1,s\shuffle\{2,\cdots,n-1\},n|\sigma_{n+1})=0\,,~~\label{bcj}
\eea
since it will be used subsequently. The combinatory momentum $Y_s$ is defined as the summation of external momenta carried by legs at the l.h.s of $s$ in the ordering $(1,s\shuffle\{2,\cdots,n-1\},n)$, the ordering $\sigma_{n+1}$ is defined for $n+1$ legs in $\{1,\cdots,n\}\cup s$.
BCJ relations imply the independence of BAS amplitudes in KK basis, thus the so called BCJ basis can be chosen as BAS amplitudes with three fixed legs in color orderings. However, in BCJ relations, coefficients of BAS amplitudes depend on Mandelstam variables, this character leads to poles in coefficients when expanding to BCJ basis. On the other hand, when expanding to KK basis, coefficients contain no pole. In this note, we choose the KK basis since we expect that all poles are included in basis, and coefficients only contribute numerators.

\subsection{Recursive expansion of single-trace YMS amplitudes}
\label{subsec-expan-YMS}

The YMS theory under consideration is the massless ${\rm YM}\oplus {\rm BAS}$ theory with Lagrangian \cite{Chiodaroli:2017ngp}
\bea
{\cal L}_{{\rm YM}\oplus {\rm BAS}}&=&-{1\over4}\,F^{a}_{\mu\nu}\,F^{a\mu\nu}+{1\over2}\,D_\mu\phi^{Aa}\,D^{\mu}\phi^{Aa}
-{g^2\over4}\,f^{abe}\,f^{ecd}\,\phi^{Aa}\phi^{Bb}\phi^{Ac}\phi^{Bd}\nn
& &+{\lambda \over3!}\,F^{ABC}f^{abc}\,
\phi^{Aa}\phi^{Bb}\phi^{Cc}\,.~~\label{lag}
\eea
The indices $a$, $b$, $c$ and $d$ run over the adjoint representation of the gauge group. Scalar fields carry additional flavor indices $A$, $B$, $C$. The field strength and covariant derivative are defined in the usual way
\bea
& &F^{a}_{\mu\nu}=\partial_\mu\,A^a_\nu-\partial_\nu\,A^a_\mu+g\,f^{abc}\,A^b_\mu A^c_\nu\,,\nn
& &D_\mu\phi^{Aa}=\partial_\mu\phi^{Aa}+g\,f^{abc}\,A^b_\mu\phi^{Ac}\,.
\eea
The general tree amplitude of this theory includes both massless external scalars and massless external gluons.
Through the standard technic, one can decompose the gauge group factors to obtain
\bea
{\cal A}_{n+m}=\sum_{\sigma\in{\cal S}_{n+m}}\,{\rm Tr}[T^{a_{\sigma_1}}\cdots T^{a_{\sigma_{n+m}}}]\,{\cal A}_{n+m,1}(\sigma_1,\cdots,\sigma_{n+m})\,,~~\label{deco-color}
\eea
where ${s}_{n+m}$ denotes non-cyclic permutations among $n$ external scalars and $m$ external gluons, $T^{a_{\sigma_i}}$ encodes the generator of the gauge group. Here ${\cal A}_{n+m}$ is the kinematic part of the amplitude without coupling constants. Meanwhile, decomposition of the flavor group factors leads to the structure of tree amplitudes which is similar to that of loop amplitudes in that, unlike pure gauge theories, it is not restricted to have only single-trace term \cite{Chiodaroli:2014xia,Chiodaroli:2017ngp}:
\bea
{\cal A}_{n+m}&=&\sum_{\sigma\in{\cal S}_n}\,{\rm Tr}[T^{A_{\sigma_1}}\cdots T^{A_{\sigma_m}}]\,{\cal A}_{n+m,1}(\sigma_1,\cdots,\sigma_n)\nn
& &+\sum_{n_1+n_2=n}\,\sum_{\a\in{\cal S}_{n_1}}\,\sum_{\b\in{\cal S}_{n_2}}\,{\rm Tr}[T^{A_{\a_1}}\cdots T^{A_{\a_{n_1}}}]\,{\rm Tr}[T^{A_{\b_1}}\cdots T^{A_{\b_{n_2}}}]\,{\cal A}_{n+m,2}(\a_1,\cdots,\a_{n_1}|\b_1,\cdots,\b_{n_2})\nn
& &+\cdots\,,~~\label{multi-trace}
\eea
where ${\cal S}_{n_i}$ again stands for the set of non-cyclic permutations. The single-trace and double-trace terms are collected in the first and second lines respectively, and $\cdots$ in the third line denotes the remaining multi-trace terms. External scalars for the partial amplitudes ${\cal A}_{n+m,2}(\a_1,\cdots,\a_{n_1}|\b_1,\cdots,\b_{n_2})$ in the second line are grouped into two orderings which are $(\a_1,\cdots,\a_{n_1})$
and $(\b_1,\cdots,\b_{n_2})$. Analogously, partial amplitudes in multi-trace terms in the third line include more orderings among external scalars.

In this note, we focus on single-trace partial tree amplitudes ${\cal A}_{n+m,1}(\sigma_1,\cdots,\sigma_n)$ in the first line of \eref{multi-trace}, due to the following reasons. First, there are special connections between tree YM amplitudes, tree BAS amplitudes, and single-trace tree YMS amplitudes \cite{Cachazo:2013iea,Cachazo:2014xea}. These relations can be extended to tree GR amplitudes, tree YM amplitudes, and single-trace tree EYM amplitudes, via the well known double copy structure \cite{Cachazo:2013iea,Cachazo:2014nsa,Cachazo:2014xea,Zhou:2018wvn,Bollmann:2018edb}.
Secondly, single-trace tree YMS amplitudes play the crucial role  when studying color-kinematic duality \cite{Cachazo:2013iea,Cachazo:2014nsa,Cachazo:2014xea,Cheung:2021zvb,Chiodaroli:2014xia,Chiodaroli:2017ngp}. By definition, single-trace YMS amplitudes under consideration are partial amplitudes ${\cal A}_{\rm YS}(\sigma_n;\{p_i\}_m|\sigma_{n+m})$, obtained by decomposing flavor and gauge group factors simultaneously. Here the color ordering in \eref{deco-color} among all external legs is denoted as $\sigma_{n+m}$, while the flavor ordering among external scalars, for single-trace terms in the first line in \eref{multi-trace}, is labeled as $\sigma_n$. The un-ordered set of $m$ external gluons is denoted by $\{p_i\}_m\equiv\{p_1,\cdots,p_m\}$. In other words, gluons belong to only one ordering $\sigma_{n+m}$.
Notice that the single-trace sector of YMS theory is equivalent to dropping the $\phi^4$ terms in the Lagrangian \eref{lag}, see in \cite{Cheung:2021zvb,Chiodaroli:2014xia,Chiodaroli:2017ngp}.

The discussion for expansions of tree level amplitudes in the previous subsection.\ref{subsecexpand} indicates that the single-trace tree YMS amplitude ${\cal A}_{\rm YS}(1,\cdots,n;\{p_i\}_m|1,\sigma_{n+m-2},n)$ can be expanded to KK BAS basis as
\bea
{\cal A}_{\rm YS}(1,\cdots,n;\{p_i\}_m|1,\sigma_{n+m-2},n)
&=&\sum_{\W\sigma_{n+m-2}}\,{\cal C}(\W\sigma_{n+m-2},\epsilon_i,k_j)\,{\cal A}_{\rm S}(1,\W\sigma_{n+m-2},n|1,\sigma_{n+m-2},n)\,,~~~\label{exp-YMS-KK}
\eea
where $\sigma_{n+m-2}$ and $\W\sigma_{n+m-2}$ denote orderings among external legs in $\{2,\cdots,n-1\}\cup\{p_i\}_m$. The double copy structure \cite{Kawai:1985xq,Bern:2008qj,Chiodaroli:2014xia,Johansson:2015oia,Johansson:2019dnu,Cachazo:2013iea} indicates that the coefficient ${\cal C}(\W\sigma_{n+m-2},\epsilon_i,k_j)$ depends on polarization vectors $\epsilon_i$ carried by external gluons,
momenta $k_i$ carried by either gluons or scalars, and orderings $\W\sigma_{n+m-2}$, but is independent of the ordering $\sigma_{n+m-2}$\footnote{Originally, the double copy means the GR amplitude can be factorized as ${\cal A}_{\rm G}={\cal A}_{\rm YM}\times {\cal S}\times{\cal A}_{\rm YM}$, where the kernel ${\cal S}$ is obtained by inverting BAS amplitudes. Our assumption that the coefficients depend on only one color ordering is equivalent to the original version, see in \cite{Zhou:2022orv}.}. The independence of the ordering $\sigma_{n+m-2}$ leads to the more general ansatz
\bea
{\cal A}_{\rm YS}(1,\cdots,n;\{p_i\}_m|\sigma_{n+m})
&=&\sum_{\W\sigma_{n+m-2}}\,{\cal C}(\W\sigma_{n+m-2},\epsilon_i,k_j)\,{\cal A}_{\rm S}(1,\W\sigma_{n+m-2},n|\sigma_{n+m})\,,~~~\label{exp-YS-KK}
\eea
where $\sigma_{n+m}$ stands for the general ordering among all external legs, without fixing any one at any particular position.

The expansion in \eref{exp-YS-KK} can be achieved by applying the following recursive expansion iteratively \cite{Fu:2017uzt,Teng:2017tbo,Feng:2019tvb},
\bea
& &{\cal A}_{\rm YS}(1,\cdots,n;\{p_i\}_m|\sigma_{n+m})\nn
&=&\sum_{\vec{\pmb{\a}}}\,\big(\epsilon_p\cdot F_{\vec{\pmb{\a}}}\cdot Y_{\vec{\pmb{\a}}}\big)\,{\cal A}_{\rm YS}(1,\{2,\cdots,n-1\}\shuffle \{\vec{\pmb{\a}},p\},n;\{p_i\}_m\setminus\{p\cup\pmb{\a}\}|\sigma_{n+m})\,,~~\label{expan-YMS-recur}
\eea
where $p$ is the fiducial gluon which can be chosen as any element in $\{p_i\}_m$, and $\pmb{\a}$ are subsets of $\{p_i\}_m\setminus p$ which is allowed to be empty. When $\pmb{\a}=\{p_i\}_m\setminus p$, the YMS amplitudes in the second line of \eref{expan-YMS-recur} are reduced to pure BAS ones. The ordered set $\vec{\pmb{\a}}$ is generated from $\pmb{\a}$
by endowing an order among elements in $\pmb{\a}$. The tensor $F^{\mu\nu}_{\vec{\pmb{\a}}}$ is defined as
\bea
F^{\mu\nu}_{\vec{\pmb{\a}}}\equiv\big(f_{\a_k}\cdot f_{\a_{k-1}}\cdots f_{\a_2}\cdot f_{\a_1}\big)^{\mu\nu}\,,
\eea
for $\vec{\pmb{\a}}=\{\a_1,\cdots\a_k\}$, where each anti-symmetric strength tensor $f_i$ is given as $f^{\mu\nu}_i\equiv k^\mu_i\epsilon^\nu_i-\epsilon^\mu_i k^\nu_i$. The combinatory momentum $Y_{\vec{\pmb{\a}}}$ is the summation of momenta carried by external scalars at the l.h.s of $\a_1$ in the color ordering $(1,\{2,\cdots,n-1\}\shuffle \vec{\pmb{\a}},n)$, where $\a_1$ is the first element in the ordered set $\vec{\pmb{\a}}$. The symbol $\shuffle$ means summing over permutations of $\{2,\cdots,n-1\}\cup\{\vec{\pmb{\a}},p\}$ which preserve the orderings of two ordered sets $\{2,\cdots,n-1\}$ and $\{\vec{\pmb{\a}},p\}$, as explained around \eref{shuffle}. The summation in \eref{expan-YMS-recur} is over all un-equivalent ordered sets $\vec{\pmb{\a}}$. In the recursive expansion \eref{expan-YMS-recur}, the YMS amplitude is expanded to YMS amplitudes with less gluons and more scalars. Repeating such expansion, one can finally expand any YMS amplitude to pure BAS ones.

In the recursive expansion \eref{expan-YMS-recur}, the gauge invariance for each gluon in $\{p_i\}_m\setminus p$ is manifest, since the tensor $f^{\mu\nu}$ vanishes automatically under the replacement $\epsilon_i\to k_i$, due to the definition. However, the gauge invariance for the fiducial gluon $p$ has not been manifested. When applying \eref{expan-YMS-recur} iteratively, a fiducial gluon will be required at every step. Consequently, in the resulted expansion to pure BAS amplitudes, the gauge invariance for each polarization will be spoiled. Furthermore, the manifest permutation invariance among external gluons are also broken. Notice that the breaking of manifest permutation invariance among external scalars can not be avoided, since the KK basis requires fixing two legs at two special positions in orderings. However, a special external gluon is not necessary. To obtain the expansion which manifests the gauge invariance and permutation invariance for external gluons simultaneously, one should employ another recursive expansion
\bea
& &{\cal A}_{\rm YS}(1,\cdots,n;\{p_i\}_m|\sigma_{n+m})\nn
&=&\sum_{\vec{\pmb{\a}}}\,{k_r\cdot F_{\vec{\pmb{\a}}}\cdot Y_{\vec{\pmb{\a}}}\over k_r\cdot k_{p_1\cdots p_m}}\,{\cal A}_{\rm YS}(1,\{2,\cdots,n-1\}\shuffle \vec{\pmb{\a}},n;\{p_i\}_m\setminus\pmb{\a}|\sigma_{n+m})\,,~~\label{expan-YMS-gi}
\eea
where $k_r$ is a reference massless momentum. Here the notations are parallel to those for \eref{expan-YMS-recur}. The formula \eref{expan-YMS-gi} was found by Clifford Cheung and James Mangan, in the so called covariant color-kinematic duality framework \cite{Cheung:2021zvb}. The expansion \eref{expan-YMS-gi} does not require any fiducial gluon, the gauge invariance for any polarization and the permutation symmetry among external gluons are manifested. Using \eref{expan-YMS-recur} iteratively, one finally obtains the new expansion of YMS amplitudes to KK BAS basis, with explicit gauge and permutation invariance for gluons. Reconstruct the expansion \eref{expan-YMS-gi} through a bottom up method is the main purpose of this note.

\subsection{Soft theorems for external scalars and gluons}
\label{subsec-soft}

In this subsection, we rapidly review the soft theorems for external scalars and gluons, which are crucial for subsequent constructions in next section.

For the double color ordered BAS amplitude ${\cal A}_{\rm S}(1,\cdots,n|\sigma_n)$, we re-scale $k_i$
as $k_i\to\tau k_i$, and expand the amplitude by $\tau$. The leading order contribution aries from propagators $1/s_{1(i+1)}$
and $1/ s_{(i-1)i}$ which are at the $\tau^{-1}$ order,
\bea
{\cal A}^{(0)_i}_{\rm S}(1,\cdots,n|\sigma_n)&=&{1\over \tau}\Big({\delta_{i(i+1)}\over s_{i(i+1)}}+{\delta_{(i-1)i}\over s_{(i-1)i}}\Big)\,
{\cal A}_{\rm S}(1,\cdots,i-1,\not{i},i+1,\cdots,n|\sigma_n\setminus i)\nn
&=&S^{(0)_i}_s\,{\cal A}_{\rm S}(1,\cdots,i-1,\not{i},i+1,\cdots,n|\sigma_n\setminus i)\,,~~~\label{soft-theo-s}
\eea
where $\not{i}$ stands for removing the leg $i$, $\sigma_n\setminus i$ means the color ordering generated from $\sigma_n$ by eliminating $i$. The superscript $(0)_i$ is introduced for denoting the leading order when considering the soft behaviour of $k_i$. The symbol $\delta_{ab}$ is defined
as follows\footnote{The Kronecker symbol will not appear in this paper, thus we hope the notation $\delta_{ab}$ will not confuse the readers.} \cite{Zhou:2022orv}. In an ordering, if two legs $a$ and $b$ are adjacent, then $\delta_{ab}=1$ if $a$ precedes $b$, and $\delta_{ab}=-1$ if $a$ follows $b$. If $a$ and $b$ are not adjacent, $\delta_{ab}=0$. From the definition, it is straightforward to observe $\delta_{ab}=-\delta_{ba}$. In \eref{soft-theo-s}, $\delta_{i(i+1)}$ and $\delta_{(i-1)i}$ are defined for the ordering $\sigma_n$, and the leading soft operator $S^{(0)_i}_s$ for the scalar $i$ is given as 
\bea
S^{(0)_i}_s={1\over \tau}\,\Big({\delta_{i(i+1)}\over s_{i(i+1)}}+{\delta_{(i-1)i}\over s_{(i-1)i}}\Big)\,.~~~~\label{soft-fac-s-0}
\eea
By requiring the universality of soft behavior, the above result can be generated to the YMS amplitude as \cite{Zhou:2022orv}
\bea
{\cal A}^{(0)_i}_{\rm YS}(1,\cdots,n;\{p_i\}_m|\sigma_{n+m})
&=&S^{(0)_i}_s\,{\cal A}_{\rm YS}(1,\cdots,i-1,\not{i},i+1,\cdots,n;\{p_i\}_m|\sigma_{n+m}\setminus i)\,,~~~\label{soft-theo-s2}
\eea
where the soft factor $S^{(0)_i}_s$ is the same as in \eref{soft-fac-s-0}. In other words, the soft operator $S^{(0)_i}_s$ does not act on any external gluon.

The soft theorems for external gluons at leading and sub-leading orders can be obtained via various approaches \cite{Casali:2014xpa,Schwab:2014xua}. Notice that one of these methods is to use the expanded formula of YMS amplitude in \eref{expan-YMS-recur}, as can be seen in \cite{Zhou:2022orv}. Such soft theorems are given as
\bea
{\cal A}^{(0)_{p_j}}_{\rm YS}(1,\cdots,n;\{p_i\}_m|\sigma_{n+m})
&=&S^{(0)_{p_j}}_g\,{\cal A}_{\rm YS}(1,\cdots,n;\{p_i\}_m\setminus p_j|\sigma_{n+m}\setminus p_j)\,,~~~\label{soft-theo-g0}
\eea
and
\bea
{\cal A}^{(1)_{p_j}}_{\rm YS}(1,\cdots,n;\{p_i\}_m|\sigma_{n+m})
&=&S^{(1)_{p_j}}_g\,{\cal A}_{\rm YS}(1,\cdots,n;\{p_i\}_m\setminus p_j|\sigma_{n+m}\setminus p_i)\,,~~~\label{soft-theo-g1}
\eea
where the external momentum $k_{p_j}$ is re-scaled as $k_{p_j}\to\tau k_{p_j}$. The soft factors at leading and sub-leading orders are given by
\bea
S^{(0)_{p_j}}_g={1\over\tau}\,\sum_{a\in\{1,\cdots,n\}\cup\{p_i\}_m\setminus p_j}\,{\delta_{ap_j}\,(\epsilon_{p_j}\cdot k_a)\over s_{ap_j}}\,,~~\label{soft-fac-g-0}
\eea
and
\bea
S^{(1)_{p_j}}_g=\sum_{a\in\{1,\cdots,n\}\cup\{p_i\}_m\setminus p_j}\,{\delta_{ap_j}\,\big(\epsilon_{p_j}\cdot J_a\cdot k_{p_j}\big)\over s_{ap_j}}\,,~~\label{soft-fac-g-1}
\eea
respectively. In \eref{soft-fac-g-0} and \eref{soft-fac-g-1}, one should sum over all external legs $a$, i.e., these soft operators for external gluon act on both external scalars and gluons.

The sub-leading soft operator \eref{soft-fac-g-1} for external gluon plays the central role in the next section. Here we list some useful results for the action of this operator. The angular momentum operator $J_a^{\mu\nu}$ acts on Lorentz vector $k^\rho_a$ with the orbital part of the generator, and on $\epsilon^\rho_a$ with the spin part of the generator in the vector representation,
\bea
J_a^{\mu\nu}\,k_a^\rho=k_a^{[\mu}\,{\partial k_a^\rho\over\partial k_{a,\nu]}}\,,~~~~
J_a^{\mu\nu}\,\epsilon_a^\rho=\big(\eta^{\nu\rho}\,\delta^\mu_\sigma-\eta^{\mu\rho}\,\delta^\nu_\sigma\big)\,\epsilon^\sigma_a\,.
\eea
Then the action of $S^{(1)_p}_g$ can be re-expressed as
\bea
S^{(1)_p}_g&=&-\sum_{V_a}\,{\delta_{ap}\over s_{ap}}\,V_a\cdot f_p\cdot{\partial\over\partial V_a}\,,~~\label{act-s-V}
\eea
due to the observation that the amplitude is linear in each polarization vector. In \eref{act-s-V}, the summation over $V_a$
is among all Lorentz vectors including both momenta and polarizations. The operator \eref{act-s-V} is a differential operator which satisfies Leibnitz's rule. Using \eref{act-s-V}, we immediately get
\bea
\big(S^{(1)_p}_g\,k_a\big)\cdot V=-{\delta_{ap}\over s_{ap}}\,(k_a\cdot f_p\cdot V)\,,~~~~\big(S^{(1)_p}_g\,\epsilon_a\big)\cdot V=-{\delta_{ap}\over s_{ap}}\,(\epsilon_a\cdot f_p\cdot V)\,,~~~~\label{iden-1}
\eea
where $V$ is an arbitrary Lorentz vector, and
\bea
V_1\cdot\big(S^{(1)_p}_g\,f_a\big)\cdot V_2={\delta_{ap}\over s_{ap}}\,V_1\cdot(f_p\cdot f_a-f_a\cdot f_p)\cdot V_2\,,~~~~\label{iden-2}
\eea
for two arbitrary Lorentz vectors $V_1$ and $V_2$, where the anti-symmetric tensor $f_i$ is defined as $f_i^{\mu\nu}\equiv k_i^\mu\epsilon_i^\nu-\epsilon_i^\mu k_i^\nu$, as introduced previously.

\section{Expansion of single-trace YMS amplitude}
\label{sec-expan-YMS}

In this section, we study the expansion of single-trace tree YMS amplitudes.
Our purpose is to reconstruct the expansion in \eref{expan-YMS-gi} through a purely bottom up approach. For simplicity, we will chose the reference momentum $k_r$ in \eref{expan-YMS-gi} as $k_r=k_n$. We first bootstrap the $3$-point YMS amplitudes with only one external gluon. Then, we invert the soft theorem for BAS scalars to construct YMS amplitudes with more external scalars, while keeping the number of external gluon to be fixed. After this construction, we use the BCJ relation to transmute the resulted expansion of such special BAS amplitudes with only one external gluon to a manifestly gauge invariant form. Next, we invert the sub-leading soft theorem for external gluons, to give a recursive pattern which leads to the general expansion of single-trace YMS amplitudes with arbitrary number of external gluons,
which is manifestly gauge invariant for any polarization. Through the whole process, the manifest permutation symmetry among external gluons is kept at each step.

\subsection{YMS amplitude with one external gluon}
\label{subsec-1gluon}

In this subsection, we construct the single-trace YMS amplitude ${\cal A}_{\rm YS}(1,\cdots,n;p|\sigma_{n+1})$ which contains external scalars $i\in\{1,\cdots,n\}$ and only one external gluon $p$, by using the purely bottom up method. To start, we first bootstrap $3$-point amplitudes ${\cal A}_{\rm YS}(1,2;p|\sigma_3)$.
In $d$-dimensional space-time, any $n$-point amplitude has the mass dimension $d-{d-2\over2}n$, thus the $3$-point one has the mass dimension $3-{d\over2}$. On the other hand, the coupling constant $g$ in Lagrangian \eref{lag} has mass dimension $2-{d\over2}$, thus the kinematic part ${\cal A}_{\rm YS}(1,2;p|\sigma_3)$ has mass dimension $1$. Meanwhile, the amplitude ${\cal A}_{\rm YS}(1,2;p|\sigma_3)$ with one external gluon $p$
should be linear in $\epsilon_p$, where $\epsilon_p$ is the polarization vector carried by $p$. Finally, the $3$-point amplitude does not include any pole since it can never be factorized into lower-point amplitudes. The above constraints uniquely fix ${\cal A}_{\rm YS}(1,2;p|\sigma_3)$ to be $\epsilon_p\cdot k_1$, up to an overall sign\footnote{When talking about $3$-point amplitudes, we allow the components of external momenta to take complex values, otherwise the momentum conservation and on-shell conditions can not be satisfied simultaneously. A well known example is the MHV and anti-MHV YM amplitudes in spinor-helicity representation in $4$-d space-time.}. Notice that $\epsilon_p\cdot k_2$ is equivalent to $\epsilon_p\cdot k_1$, due to the momentum conservation and the on-shell condition $\epsilon_p\cdot k_p=0$. Using the observation ${\cal A}_{\rm S}(1,p,2|1,p,2)=1$, we arrive at the following expansion
\bea
{\cal A}_{\rm YS}(1,2;p|\sigma_3)=(\epsilon_p\cdot k_1)\,{\cal A}_{\rm S}(1,p,2|\sigma_3)\,,~~\label{start-point}
\eea
coincides with the general ansatz in \eref{exp-YS-KK}.
Here the overall sign of ${\cal A}_{\rm YS}(1,2;p|1,p,2)$ is chosen to be $1$, which forces the overall sign of ${\cal A}_{\rm YS}(1,2;p|2,p,1)$
to be $-1$. Notice that $\sigma_3$ has only two inequivalent choices which are $(1,p,2)$ and $(2,p,1)$, due to the cyclic symmetry of ordering.

Then we invert the leading soft theorem for external scalars in \eref{soft-theo-s} and \eref{soft-fac-s-0}, to insert external scalars into ${\cal A}_{\rm YS}(1,2;p|2,p,1)$. Consider the $4$-point amplitude ${\cal A}_{\rm YS}(1,2,3;p|\sigma_4)$ with external scalars $i\in\{1,2,3\}$ and external gluon $p$. The soft theorem in \eref{soft-theo-s} and \eref{soft-fac-s-0} indicates the following leading soft behavior when taking $k_2\to\tau k_2$,
\bea
{\cal A}_{\rm YS}^{(0)_2}(1,2,3;p|\sigma_4)&=&S^{(0)_2}_s\,{\cal A}_{\rm YS}^{(0)_2}(1,3;p|\sigma_4\setminus2)\nn
&=&\Big({\delta_{12}\over s_{12}}+{\delta_{23}\over s_{23}}\Big)\,\Big((\epsilon_p\cdot k_1)\,{\cal A}_{\rm S}(1,p,3|\sigma_4\setminus2)\Big)\nn
&=&\Big({\delta_{12}\over s_{12}}+{\delta_{2p}\over s_{2p}}+{\delta_{p2}\over s_{p2}}+{\delta_{23}\over s_{23}}\Big)\,\Big((\epsilon_p\cdot k_1)\,{\cal A}_{\rm S}(1,p,3|\sigma_4\setminus2)\Big)\nn
&=&(\epsilon_p\cdot k_1)\,\Big({\cal A}^{(0)_2}_{\rm S}(1,2,p,3|\sigma_4)+{\cal A}^{(0)_2}_{\rm S}(1,p,2,3|\sigma_4)\Big)\,,~~\label{4s-2soft}
\eea
where the second equality uses the soft factor in \eref{soft-fac-s-0} and the expansion \eref{start-point} simultaneously, the third uses the anti-symmetry of $\delta_{ab}$, while the third uses the inversion of soft theorem in \eref{soft-theo-s} and \eref{soft-fac-s-0}.
The soft behavior in \eref{4s-2soft} implies that ${\cal A}_{\rm YS}(1,2,3;p|\sigma_4)$ can be expanded as
\bea
{\cal A}_{\rm YS}(1,2,3;p|\sigma_4)&=&C(1,2,p,3)\,{\cal A}_{\rm S}(1,2,p,3|\sigma_4)\nn
& &+C(1,p,2,3)\,{\cal A}_{\rm S}(1,p,2,3|\sigma_4)\,,
\eea
where coefficients $C(1,2,p,3)$ and $C(1,p,2,3)$ satisfy
\bea
C^{(0)_2}(1,2,p,3)=C^{(0)_2}(1,p,2,3)=\epsilon_p\cdot k_1\,.~~\label{constraint}
\eea
The constraint \eref{constraint} requires $C(1,2,p,3)$ and $C(1,p,2,3)$ to have the form
\bea
C(1,2,p,3)=\epsilon_p\cdot k_1+\a_1\,\epsilon_p\cdot k_2\,,~~~~C(1,p,2,3)=\epsilon_p\cdot k_1+\a_2\,\epsilon_p\cdot k_2\,.
\eea
To determine $\a_1$ and $\a_2$, one can consider the leading soft behavior of $k_1\to\tau k_1$ to obtain
\bea
{\cal A}^{(0)_1}_{\rm YS}(1,2,3;p|\sigma_4)&=&S^{(0)_1}_s\,{\cal A}_{\rm YS}(2,3;p|\sigma_4\setminus1)\nn
&=&\Big({\delta_{31}\over s_{31}}+{\delta_{12}\over s_{12}}\Big)\,\Big((\epsilon_p\cdot k_2)\,{\cal A}_{\rm S}(2,p,3|\sigma_4\setminus1)\Big)\nn
&=&-\Big({\delta_{21}\over s_{21}}+{\delta_{1p}\over s_{1p}}+{\delta_{p1}\over s_{1p}}+{\delta_{13}\over s_{13}}\Big)\,\Big((\epsilon_p\cdot k_2)\,{\cal A}_{\rm S}(2,p,3|\sigma_4\setminus1)\Big)\nn
&=&-(\epsilon_p\cdot k_2)\,\Big({\cal A}^{(0)_1}_{\rm S}(2,1,p,3|\sigma_4\setminus1)+{\cal A}^{(0)_1}_{\rm S}(2,p,1,3|\sigma_4\setminus1)\Big)\nn
&=&(\epsilon_p\cdot k_2)\,{\cal A}^{(0)_1}_{\rm S}(1,2,p,3|\sigma_4\setminus1)\,,~~\label{4p-1soft}
\eea
where the last equality uses the fundamental BCJ relation in \eref{bcj}. The soft behavior in \eref{4p-1soft} indicates
\bea
C^{(0)_1}(1,2,p,3)=\epsilon_p\cdot k_2\,,~~~~C^{(0)_1}(1,p,2,3)=0\,.~~\label{constraint2}
\eea
Comparing \eref{constraint2} with \eref{constraint}, we find $\a_1=1$, $\a_2=0$, thus the full expansion of ${\cal A}_{\rm YS}(1,2,3;p|\sigma_4)$ is given as
\bea
{\cal A}_{\rm YS}(1,2,3;p|\sigma_4)&=&(\epsilon_{p}\cdot k_{12})\,{\cal A}_{\rm S}(1,2,p,3|\sigma_4)+(\epsilon_p\cdot k_1)\,{\cal A}_{\rm S}(1,p,2,3|\sigma_4)\nn
&=&(\epsilon_p\cdot Y_p)\,{\cal A}_{\rm S}(1,2\shuffle p,3|\sigma_4)\,,
\eea
where the symbol $\shuffle$ is defined around \eref{shuffle}, and the combinatory momentum $Y_p$ is defined as the summation over momenta carried by scalars at the l.h.s of $p$ in the color ordering $(1,2\shuffle p,3)$.

Repeating the above manipulation, one can determine the general
YMS amplitude ${\cal A}_{\rm YS}(1,\cdots,n;p|\sigma_{n+1})$ with arbitrary number of external scalars and one external gluon $p$, in the expanded formula
\bea
{\cal A}_{\rm YS}(1,\cdots,n;p|\sigma_{n+1})=(\epsilon_p\cdot Y_p)\,{\cal A}_{\rm S}(1,\{2,\cdots,n-1\}\shuffle p,n|\sigma_{n+1})\,,~~~~\label{expan-1g-old}
\eea
where the combinatory momentum $Y_p$ was defined in analogously.
In the above expansion, it is direct to observe the gauge invariance for the polarization $\epsilon_p$, since the replacement $\epsilon_p^\mu\to k_p^\mu$ yields the BCJ relation \eref{bcj},
\bea
0=(k_p\cdot Y_p)\,{\cal A}_{\rm S}(1,\{2,\cdots,n-1\}\shuffle p,n|\sigma_{n+1})\,.
\eea
However, if one use the formula \eref{expan-1g-old} and the soft theorem for external gluons to construct YMS amplitudes with more external gluons, the resulting expansion is in the formula \eref{expan-YMS-recur} \cite{Zhou:2022orv}, and
the gauge invariance for the polarization of the fiducial gluon is obscure. The proof of gauge invariance for such fiducial polarization requires the application of BCJ relations in a complicated way, and the complexity increases rapidly as the number of external gluons increases. To avoid this disadvantage, we want to
rewrite \eref{expan-1g-old} so that the manifest gauge invariance is carried by coefficients. To realize the goal,
one can use the BCJ relation to modify \eref{expan-1g-old} as
\bea
{\cal A}_{\rm YS}(1,\cdots,n;p|\sigma_{n+1})
&=&(\epsilon_p\cdot Y_p)\,{\cal A}_{\rm S}(1,\{2,\cdots,n-1\}\shuffle p,n|\sigma_{n+1})\nn
& &-{k_n\cdot\epsilon_p\over k_n\cdot k_p}\,(k_p\cdot Y_p)\,{\cal A}_{\rm S}(1,\{2,\cdots,n-1\}\shuffle p,n|\sigma_{n+1})\nn
&=&{k_n\cdot f_p\cdot Y_p\over k_n\cdot k_p}\,{\cal A}_{\rm S}(1,\{2,\cdots,n-1\}\shuffle p,n|\sigma_{n+1})\,.~~~~\label{expan-1g-gi}
\eea
In the last line at the r.h.s of \eref{expan-1g-gi}, the coefficients vanish under the replacement $\epsilon_p^\mu\to k_p^\mu$,
due to the definition $f^{\mu\nu}_p\equiv k^\mu_p\epsilon^\nu_p-\epsilon^\mu_p k^\nu_p$. This is the desired new expansion for the YMS with one external gluon, trivialized the gauge invariance of polarization $\epsilon^\mu_p$, with the cost that a spurious pole $k_n\cdot k_p$ is introduced. As will be seen, by applying the recursive technic based on the soft theorem for external gluons, this new expansion leads to general expansion of YMS amplitudes with arbitrary number of external gluons, which manifests the gauge invariance for each polarization.

\subsection{YMS amplitude with two external gluons}
\label{subsec-2gluon}

Now we turn to the single-trace YMS amplitude ${\cal A}_{\rm YS}(1,\cdots,n;\{p_i\}_2|\sigma_{n+2})$ with two external gluons labeled as $p_1$ and $p_2$. Let us
consider the soft behavior of gluon $p_2$, i.e., we re-scale $k_{p_2}$ as $k_{p_2}\to\tau k_{p_2}$, and expand ${\cal A}_{\rm YS}(1,\cdots,n;\{p_i\}_2|\sigma_{n+2})$ in $\tau$. The soft theorem \eref{soft-theo-g1} indicates that the contribution at sub-leading order should be
\bea
{\cal A}^{(1)_{p_2}}_{\rm YS}(1,\cdots,n;\{p_i\}_2|\sigma_{n+2})&=&S^{(1)_{p_2}}_g\,{\cal A}_{\rm YS}(1,\cdots,n;p_1|\sigma_{n+2}\setminus p_2)\nn
&=&S^{(1)_{p_2}}_g\,\Big[{k_n\cdot f_{p_1}\cdot Y_{p_1}\over k_n\cdot k_{p_1}}\,{\cal A}_{\rm S}(1,\{2,\cdots,n-1\}\shuffle p_1,n|\sigma_{n+2}\setminus p_2)\Big]\nn
&=&B_1+B_2+B_3\,,~~~\label{2g-Lb}
\eea
where
\bea
B_1={k_n\cdot f_{p_1}\cdot Y_{p_1}\over k_n\cdot k_{p_1}}\,{\cal A}^{(1)_{p_2}}_{\rm YS}(1,\{2,\cdots,n-1\}\shuffle p_1,n;p_2|\sigma_{n+2})\,,~~\label{B1}
\eea
\bea
B_2&=&\tau\,{k_n\cdot f_{p_2}\cdot f_{p_1}\cdot Y_{p_1}\over k_n\cdot k_{p_1}}\,{\cal A}^{(0)_{p_2}}_{\rm S}(1,\{2,\cdots,n-1\}\shuffle \{p_1,p_2\},n|\sigma_{n+2})\nn
& &+\tau\,{k_n\cdot f_{p_1}\cdot f_{p_2}\cdot Y_{p_2}\over k_n\cdot k_{p_1}}\,{\cal A}^{(0)_{p_2}}_{\rm S}(1,\{2,\cdots,n-1\}\shuffle \{p_2,p_1\},n|\sigma_{n+2})\,,~~\label{B2}
\eea
and
\bea
B_3=-\tau\,{(k_n\cdot f_{p_1}\cdot Y_{p_1})\,(k_n\cdot f_{p_2}\cdot k_{p_1})\over (k_n\cdot k_{p_1})^2}\,{\cal A}^{(0)_{p_2}}_{\rm S}(1,\{2,\cdots,n-1\}\shuffle \{p_1,p_2\},n|\sigma_{n+2})\,.~~~\label{B3}
\eea
The second equality of \eref{2g-Lb} is obtained by substituting the expansion \eref{expan-1g-gi}. Three parts $B_1$, $B_2$ and $B_3$ arise from
acting $S^{(1)_{p_2}}_g$ on ${\cal A}_{\rm S}(1,\{2,\cdots,n-1\}\shuffle p_1,n|\sigma_{n+2}\setminus p_2)$, $k_n\cdot f_{p_1}\cdot Y_{p_1}$ and $k_n\cdot k_{p_1}$ respectively, due to Leibnitz's rule. The expression for the first part $B_1$ in \eref{B1} is obtained via the soft theorem
\bea
S^{(1)_{p_2}}_g\,{\cal A}_{\rm S}(1,\{2,\cdots,n-1\}\shuffle p_1,n|\sigma_{n+2}\setminus p_2)={\cal A}^{(1)_{p_2}}_{\rm YS}(1,\{2,\cdots,n-1\}\shuffle p_1,n;p_2|\sigma_{n+2})\,.
\eea
The second part $B_2$ can be calculated as follows.
Using relations \eref{iden-1} and \eref{iden-2}, we have
\bea
S^{(1)_{p_2}}_g\,(k_n\cdot f_{p_1}\cdot Y_{p_1})=H_1+H_2\,,
\eea
where
\bea
H_1&=&-{\delta_{np_2}\,(k_n\cdot f_{p_2}\cdot f_{p_1}\cdot Y_{p_1})\over s_{np_2}}
+{\delta_{p_1p_2}\,(k_n\cdot f_{p_2}\cdot f_{p_1}\cdot X_{p_1})\over s_{p_1p_2}}\nn
H_2&=&-{\delta_{p_1p_2}\,(k_n\cdot f_{p_1}\cdot f_{p_2}\cdot Y_{p_1})\over s_{p_1p_2}}
+\sum_{i=1}^j\,{\delta_{ip_2}\,(k_n\cdot f_{p_1}\cdot f_{p_2}\cdot k_i)\over s_{ip_2}}\,,~~~~\label{H1H2}
\eea
with $Y_{p_1}=\sum_{i=1}^j\,k_i$. We use $\delta_{ab}=-\delta_{ba}$ to reorganize $H_1$ as
\bea
H_1=\Big[\Big({\delta_{p_1p_2}\over s_{p_1p_2}}+{\delta_{p_2(j+1)}\over s_{p_2(j+1)}}\Big)+\sum_{k=j+1}^{n-1}\,\Big({\delta_{kp_2}\over s_{kp_2}}+{\delta_{p_2(k+1)}\over s_{p_2(k+1)}}\Big)\Big]\,(k_n\cdot f_{p_2}\cdot f_{p_1}\cdot Y_{p_1})\,.~~~~\label{h1}
\eea
The reason for expressing $H_1$ in the above manner is that we want to interpret $H_1\,{\cal A}_{\rm S}(1,\{2,\cdots,n-1\}\shuffle p_1,n|\sigma_{n+2}\setminus p_2)$ as coefficients times leading or sub-leading terms of amplitudes, as can be seen in \eref{B2}.
Combining \eref{h1} and the leading soft factor \eref{soft-fac-s-0} for the scalar, we find
\bea
& &H_1\,{\cal A}_{\rm S}(1,\{2,\cdots,n-1\}\shuffle p_1,n|\sigma_{n+2}\setminus p_2)\nn
&=&\sum_{j=1}^{n-1}\,H_1\,{\cal A}_{\rm S}(1,\cdots,j,p_1,j+1,\cdots,n|\sigma_{n+2}\setminus p_2)\nn
&=&\tau\,(k_n\cdot f_{p_2}\cdot f_{p_1}\cdot Y_{p_1})\,{\cal A}^{(0)_{p_2}}_{\rm S}(1,\{2,\cdots,n-1\}\shuffle\{p_1,p_2\},n|\sigma_{n+2})\,.~~\label{H1}
\eea
We emphasize that ${\cal A}^{(0)_{p_2}}_{\rm S}(1,\{2,\cdots,n-1\}\shuffle\{p_1,p_2\},n|\sigma_{n+2})$ is the leading order contribution of ${\cal A}_{\rm S}(1,\{2,\cdots,n-1\}\shuffle\{p_1,p_2\},n|\sigma_{n+2})$ which is at the $\tau^{-1}$ order, while $f_{p_2}$ in the coefficients is accompanied with $\tau$. Combining them together leads to the $\tau^0$ order contribution which satisfies the order of ${\cal A}^{(1)_{p_2}}_{\rm YS}(1,\cdots,n;\{p_i\}_2|\sigma_{n+2})$. Based on the reason similar as that for rewriting $H_1$, we reorganize $H_2$ as
\bea
H_2&=&\sum_{i=1}^j\,\Big({\delta_{ip_2}\over s_{ip_2}}-{\delta_{p_1p_2}\over s_{p_1p_2}}\Big)\,(k_n\cdot f_{p_1}\cdot f_{p_2}\cdot k_i)\nn
&=&\sum_{i=1}^j\,\Big[\sum_{l=i}^{j-1}\,\Big({\delta_{lp_2}\over s_{lp_2}}+{\delta_{p_2(l+1)}\over s_{p_2(l+1)}}\Big)+{\delta_{jp_2}\over s_{jp_2}}+{\delta_{p_2p_1}\over s_{p_2p_1}}\Big]\,(k_n\cdot f_{p_1}\cdot f_{p_2}\cdot k_i)\nn
&=&\sum_{l=1}^{j}\,\Big({\delta_{lp_2}\over s_{lp_2}}+{\delta_{p_2(l+1)}\over s_{p_2(l+1)}}\Big)\,\Big[k_n\cdot f_{p_1}\cdot f_{p_2}\cdot\Big(\sum_{i=1}^l\,k_i\Big)\Big]\,,
\eea
where $\delta_{q(j+1)}$ and $s_{q(j+1)}$ in the last line should be understood as $j+1=p_1$. Thus,
\bea
& &H_2\,{\cal A}_{\rm S}(1,\{2,\cdots,n-1\}\shuffle p_1,n|\sigma_{n+2}\setminus p_2)\nn
&=&\sum_{j=1}^{n-1}\,\sum_{l=1}^{j}\,\Big({\delta_{lp_2}\over s_{lp_2}}+{\delta_{p_2(l+1)}\over s_{p_2(l+1)}}\Big)\,\Big[k_n\cdot f_{p_1}\cdot f_{p_2}\cdot\Big(\sum_{i=1}^l\,k_i\Big)\Big]\,{\cal A}_{\rm S}(1,\cdots,j,p_1,j+1,\cdots,n|\sigma_{n+2}\setminus p_2)\nn
&=&\sum_{j=1}^{n-1}\,\sum_{l=1}^{j}\,\tau\,\Big[k_n\cdot f_{p_1}\cdot f_{p_2}\cdot\Big(\sum_{i=1}^l\,k_i\Big)\Big]\,{\cal A}^{(0)_{p_2}}_{\rm S}(1,\cdots,l,p_2,l+1,\cdots,j,p_1,j+1,\cdots,n|\sigma_{n+2})\nn
&=&\tau\,(k_n\cdot f_{p_1}\cdot f_{p_2}\cdot Y_{p_2})\,{\cal A}^{(0)_{p_2}}_{\rm S}(1,\{2,\cdots,n-1\}\shuffle\{p_2,p_1\},n|\sigma_{n+2})\,.~~\label{H2}
\eea
Putting \eref{H1} and \eref{H2} together yields the expression of $B_2$ in \eref{B2}. The third part $B_3$ can be calculated by using the relation \eref{iden-1},
\bea
S^{(1)_{p_2}}_g\,{1\over k_n\cdot k_{p_1}}&=&-{k_n\cdot f_{p_2}\cdot k_{p_1}\over(k_n\cdot k_{p_1})^2}\,\Big({\delta_{p_1p_2}\over s_{p_1p_2}}-{\delta_{np_2}\over s_{np_2}}\Big)\nn
&=&-{k_n\cdot f_{p_2}\cdot k_{p_1}\over(k_n\cdot k_{p_1})^2}\,\Big[{\delta_{p_1p_2}\over s_{p_1p_2}}+{\delta_{p_2a}\over s_{p_2a}}+\sum_{k=a}^{n-1}\,\Big({\delta_{kp_2}\over s_{kp_2}}+{\delta_{p_2(k+1)}\over s_{p_2(k+1)}}\Big)\Big]\,,
\eea
which indicates
\bea
& &\sum_{i=1}^{n-1}\,\Big[S^{(1)_{p_2}}_g\,{1\over k_n\cdot k_{p_1}}\Big]\,{\cal A}_{\rm S}(1,\cdots,i,p_1,i+1,\cdots,n|\sigma_{n+2}\setminus p_2)\nn
&=&-\tau\,{k_n\cdot f_{p_2}\cdot k_{p_1}\over(k_n\cdot k_{p_1})^2}\,{\cal A}^{(0)_{p_2}}_{\rm S}(1,\{2,\cdots,n-1\}\shuffle\{p_1,p_2\},n|\sigma_{n+2})\,,
\eea
therefore we get the expression of $B_3$ in \eref{B3}.

We will use $B_1$, $B_2$ and $B_3$ to reconstruct the complete expansion of ${\cal A}_{\rm YS}(1,\cdots,n;\{p_i\}_2|\sigma_{n+2})$. Mathematically, it is impossible to restore the full amplitude from the sub-leading contribution, thus further physical constrains are needed. The most important one is the symmetry under the permutation $p_1\leftrightarrow p_2$. The physical amplitude ${\cal A}_{\rm YS}(1,\cdots,n;\{p_i\}_2|\sigma_{n+2})$ should be invariant under such re-labeling. In the expansion \eref{expan-YMS-recur}, since a fiducial gluon $p_1$ is required, such symmetry is not manifest. However, one can use the BCJ relations to prove that the resulting expansions obtained by choosing different fiducial gluons are equivalent to each other. On the other hand, the purpose of the current work is to seek the new expansion which manifests the gauge invariance for all polarizations, which implies the democracy among all external gluons, thus it is natural to expect the manifested symmetry under $p_1\leftrightarrow p_2$.

With the additional condition of symmetry $p_1\leftrightarrow p_2$, let us try to find the expansion of the amplitude ${\cal A}_{\rm YS}(1,\cdots,n;\{p_i\}_2|\sigma_{n+2})$. It is straightforward to recognize that $B_2$ arises as the leading term of
\bea
P_1&=&{k_n\cdot f_{p_2}\cdot f_{p_1}\cdot Y_{p_1}\over k_n\cdot k_{p_1p_2}}\,{\cal A}_{\rm S}(1,\{2,\cdots,n-1\}\shuffle \{p_1,p_2\},n|\sigma_{n+2})\nn
& &+{k_n\cdot f_{p_1}\cdot f_{p_2}\cdot Y_{p_2}\over k_n\cdot k_{p_1p_2}}\,{\cal A}_{\rm S}(1,\{2,\cdots,n-1\}\shuffle \{p_2,p_1\},n|\sigma_{n+2})\,,~~~~\label{2g-P1}
\eea
which is at the $\tau^0$ order. The reason for turning $k_n\cdot k_{p_1}$ in the denominators to $k_n\cdot k_{p_1p_2}$ is to keep the symmetry $p_1\leftrightarrow p_2$. There is another way to keep such symmetry, which is adding terms with $k_n\cdot k_{p_2}$ in the denominator. However, ${\cal A}^{(1)_{p_2}}_{\rm YS}(1,\cdots,n;p_1,p_2|\sigma_{n+2})$ shall receive contributions from such terms, for example the coefficients carry $1/(\tau k_n\cdot k_{p_2})$ under the re-scaling $k_{p_2}\to\tau k_{p_2}$, and the amplitudes contribute $\tau^1$, which combine to terms at the $\tau^0$ order. Since such contributions do not exist in ${\cal A}^{(1)_{p_2}}_{\rm YS}(1,\cdots,n;p_1,p_2|\sigma_{n+2})$ in \eref{2g-Lb}, the $1/k_n\cdot k_{p_2}$ terms are excluded. Then the symmetry $p_1\leftrightarrow p_2$ also excludes the $1/k_n\cdot k_{p_1}$ terms, and the remaining choice is only $1/k_n\cdot k_{p_1p_2}$.

One can also observe that $B_1$ comes from the sub-leading term of
\bea
P_2={k_n\cdot f_{p_1}\cdot Y_{p_1}\over k_n\cdot k_{p_1p_2}}\,{\cal A}_{\rm YS}(1,\{2,\cdots,n-1\}\shuffle p_1,n;p_2|\sigma_{n+2})\,,~~\label{2g-P2}
\eea
which is also at the $\tau^0$ order. Again, the denominator is set to be $k_n\cdot k_{p_1p_2}$ due to the symmetry. However, the sub-leading contribution of $P_2$ receives additional contribution from expanding the denominator in $\tau$. Further more, the symmetry among two gluons requires the existence of
\bea
P_3={k_n\cdot f_{p_2}\cdot Y_{p_2}\over k_n\cdot k_{p_1p_2}}\,{\cal A}_{\rm YS}(1,\{2,\cdots,n-1\}\shuffle p_2,n;p_1|\sigma_{n+2})\,,~~\label{2g-P3}
\eea
which serves as the counter part of $P_2$ under the permutation $p_1\leftrightarrow p_2$.
These missing terms will be figured out by considering $B_3$. The expression of $B_3$ in \eref{B3} also breaks the symmetry among $p_1$ and $p_2$.
Based on such symmetry, the appearance of ${\cal A}^{(0)_{p_2}}_{\rm S}(1,\{2,\cdots,n-1\}\shuffle \{p_1,p_2\},n|\sigma_{n+2})$ in \eref{B3} implies the existence of ${\cal A}^{(0)_{p_2}}_{\rm S}(1,\{2,\cdots,n-1\}\shuffle \{p_2,p_1\},n|\sigma_{n+2})$. To realize this, we observe that
\bea
& &-(k_n\cdot f_{p_2}\cdot k_{p_1})\,{\cal A}^{(0)_{p_2}}_{\rm S}(1,\{2,\cdots,n-1\}\shuffle\{p_1,p_2\},n|\sigma_{n+2})\nn
&=&\Big(k_n\cdot f_{p_2}\cdot Y_{p_2}-k_n\cdot f_{p_2}\cdot X_{p_2} \Big)\,{\cal A}^{(0)_{p_2}}_{\rm S}(1,\{2,\cdots,n-1\}\shuffle p_1\shuffle p_2,n|\sigma_{n+2})\,,~~\label{reform-B3}
\eea
where the combinatory momentum $X_a$ is defined as the summation of momenta carried by external legs at the l.h.s of $a$ in the color ordering, without distinguishing scalars and gluons (recall that $Y_a$ only includes momenta carried by external scalars).
Then, $B_3$ can be identified as
\bea
B_3=B_4+B_5\,,
\eea
where
\bea
B_4=\tau\,{k_n\cdot f_{p_2}\cdot Y_{p_2}\over k_n\cdot k_{p_1}}\,{\cal A}^{(0)_{p_2}}_{\rm YS}(1,\{2,\cdots,n-1\}\shuffle p_2,n;p_1|\sigma_{n+2})\,,~~\label{B4}
\eea
and
\bea
B_5=-\tau\,{(k_n\cdot k_{p_2})\,(k_n\cdot f_{p_1}\cdot Y_{p_1})\over (k_n\cdot k_{p_1})^2}\,{\cal A}^{(0)_{p_2}}_{\rm YS}(1,\{2,\cdots,n-1\}\shuffle p_1,n;p_2|\sigma_{n+2})\,.~~\label{B5}
\eea
Here we have used expansions
\bea
{\cal A}_{\rm YS}(1,\{2,\cdots,n-1\}\shuffle p_2,n;p_1|\sigma_{n+2})&=&{k_n\cdot f_{p_1}\cdot Y_{p_1}\over k_n\cdot k_{p_1}}\,{\cal A}_{\rm S}(1,\{2,\cdots,n-1\}\shuffle p_1\shuffle p_2,n|\sigma_{n+2})\,,\nn
{\cal A}_{\rm YS}(1,\{2,\cdots,n-1\}\shuffle p_1,n;p_2|\sigma_{n+2})&=&{k_n\cdot f_{p_2}\cdot Y_{p_2}\over k_n\cdot k_{p_2}}\,{\cal A}_{\rm S}(1,\{2,\cdots,n-1\}\shuffle p_1\shuffle p_2,n|\sigma_{n+2})\,.~~\label{expansions}
\eea
Notice that $Y_{p_2}$ in \eref{expansions} equals to $Y_{p_2}$ in \eref{reform-B3} since in the second line of \eref{expansions} the leg $p_1$ is a scalar. The combinatory momentum $Y_{p_1}$ equals to $Y_{p_1}$ in \eref{B3} and \eref{expansions} at the leading order, since $k_{p_2}$ is accompanied with $\tau$ thus is dropped into the sub-leading term.

Now one can observe that $B_4$ is the leading term of $P_3$, and $B_5$ arises from the term at $\tau^1$ order in the expansion of the denominator of $P_2$. Thus, we conclude that
\bea
{\cal A}_{\rm YS}(1,\cdots,n;\{p_i\}_2|\sigma_{n+2})&=&P_1+P_2+P_3\nn
&=&{k_n\cdot f_{p_1}\cdot Y_{p_1}\over k_n\cdot k_{p_1p_2}}\,{\cal A}_{\rm YS}(1,\{2,\cdots,n-1\}\shuffle p_1,n;p_2|\sigma_{n+2})\nn
& &+{k_n\cdot f_{p_2}\cdot Y_{p_2}\over k_n\cdot k_{p_1p_2}}\,{\cal A}_{\rm YS}(1,\{2,\cdots,n-1\}\shuffle p_2,n;p_1|\sigma_{n+2})\nn
& &+{k_n\cdot f_{p_2}\cdot f_{p_1}\cdot Y_{p_1}\over k_n\cdot k_{p_1p_2}}\,{\cal A}_{\rm S}(1,\{2,\cdots,n-1\}\shuffle \{p_1,p_2\},n|\sigma_{n+2})\nn
& &+{k_n\cdot f_{p_1}\cdot f_{p_2}\cdot Y_{p_2}\over k_n\cdot k_{p_1p_2}}\,{\cal A}_{\rm S}(1,\{2,\cdots,n-1\}\shuffle \{p_2,p_1\},n|\sigma_{n+2})\,.~~\label{expan-2g-gi}
\eea
One can verify that taking $k_{p_1}\to\tau k_{p_1}$ and expanding \eref{expan-2g-gi} in $\tau$ give the correct behavior at sub-leading order, namely,
\bea
{\cal A}^{(1)_{p_1}}_{\rm YS}(1,\cdots,n;\{p_i\}_2|\sigma_{n+2})=S^{(1)_{p_1}}_g\,{\cal A}_{\rm YS}(1,\cdots,n;p_2|\sigma_{n+2}\setminus p_1)\,.
\eea
Notice that in the formula \eref{expan-2g-gi}, both the symmetry among two gluons and the gauge invariance for each polarization are manifested.

\subsection{YMS amplitude with three external gluons}
\label{subsec-3gluon}

The next example is the single-trace YMS amplitude ${\cal A}_{\rm YS}(1,\cdots,n;\{p_i\}_3|\sigma_{n+3})$ which includes three external gluons $p_1$, $p_2$ and $p_3$. The treatment is parallel to that in the previous subsection for the $2$-gluon case, and such similarity exhibits the recursive pattern. As in the previous case, we consider
$k_{p_3}\to\tau k_{p_3}$, and expand ${\cal A}_{\rm YS}(1,\cdots,n;\{p_i\}_3|\sigma_{n+3})$ in $\tau$. The soft theorem requires the sub-leading term to be
\bea
& &{\cal A}^{(1)_{p_3}}_{\rm YS}(1,\cdots,n;\{p_i\}_3|\sigma_{n+3})\nn
&=&S^{(1)_{p_3}}_g\,{\cal A}_{\rm YS}(1,\cdots,n;\{p_i\}_2|\sigma_{n+3}\setminus p_3)\nn
&=&S^{(1)_{p_3}}_g\,\Big[{k_n\cdot f_{p_1}\cdot Y_{p_1}\over k_n\cdot k_{p_1p_2}}\,{\cal A}_{\rm YS}(1,\{2,\cdots,n-1\}\shuffle p_1,n;p_2|\sigma_{n+3}\setminus p_3)+p_1\leftrightarrow p_2\nn
& &+{k_n\cdot f_{p_2}\cdot f_{p_1}\cdot Y_{p_1}\over k_n\cdot k_{p_1p_2}}\,{\cal A}_{\rm S}(1,\{2,\cdots,n-1\}\shuffle \{p_1,p_2\},n|\sigma_{n+3}\setminus p_3)+p_1\leftrightarrow p_2\Big]\,,
\eea
where the expanded formula \eref{expan-2g-gi} for YMS amplitude with two external gluons is used. Applying the Leibnitz's rule, we get
\bea
{\cal A}^{(1)_{p_3}}_{\rm YS}(1,\cdots,n;p_1,p_2,p_3|\sigma_{n+3})=B_1+B_2+B_3+B_4+B_5+B_6\,,
\eea
where
\bea
B_1&=&{k_n\cdot f_{p_1}\cdot Y_{p_1}\over k_n\cdot k_{p_1p_2}}\,\Big[S^{(1)_{p_3}}_g\,{\cal A}_{\rm YS}(1,\{2,\cdots,n-1\}\shuffle p_1,n;p_2|\sigma_{n+3}\setminus p_3)\Big]+p_1\leftrightarrow p_2\nn
&=&{k_n\cdot f_{p_1}\cdot Y_{p_1}\over k_n\cdot k_{p_1p_2}}\,{\cal A}^{(1)_{p_3}}_{\rm YS}(1,\{2,\cdots,n-1\}\shuffle p_1,n;\{p_i\}_3\setminus p_1|\sigma_{n+3})+p_1\leftrightarrow p_2\,,~~\label{3g-B1}
\eea
\bea
B_2&=&{k_n\cdot f_{p_2}\cdot f_{p_1}\cdot Y_{p_1}\over k_n\cdot k_{p_1p_2}}\,\Big[S^{(1)_{p_3}}_g\,{\cal A}_{\rm S}(1,\{2,\cdots,n-1\}\shuffle \{p_1,p_2\},n|\sigma_{n+3}\setminus p_3)\Big]+p_1\leftrightarrow p_2\nn
&=&{k_n\cdot f_{p_2}\cdot f_{p_1}\cdot Y_{p_1}\over k_n\cdot k_{p_1p_2}}\,{\cal A}^{(1)_{p_3}}_{\rm YS}(1,\{2,\cdots,n-1\}\shuffle \{p_1,p_2\},n;p_3|\sigma_{n+3})+p_1\leftrightarrow p_2\,,~~\label{3g-B2}
\eea
\bea
B_3&=&{\Big[S^{(1)_{p_3}}_g\,k_n\cdot f_{p_1}\cdot Y_{p_1}\Big]\over k_n\cdot k_{p_1p_2}}\,{\cal A}_{\rm YS}(1,\{2,\cdots,n-1\}\shuffle p_1,n;p_2|\sigma_{n+3}\setminus p_3)+p_1\leftrightarrow p_2\nn
&=&\tau\,{k_n\cdot f_{p_3}\cdot f_{p_1}\cdot Y_{p_1}\over k_n\cdot k_{p_1p_2}}\,{\cal A}^{(0)_{p_3}}_{\rm YS}(1,\{2,\cdots,n-1\}\shuffle \{p_1,p_3\},n;p_2|\sigma_{n+3})+p_1\leftrightarrow p_2\nn
& &+\tau\,{k_n\cdot f_{p_1}\cdot f_{p_3}\cdot Y_{p_3}\over k_n\cdot k_{p_1p_2}}\,{\cal A}^{(0)_{p_3}}_{\rm YS}(1,\{2,\cdots,n-1\}\shuffle \{p_3,p_1\},n;p_2|\sigma_{n+3})+p_1\leftrightarrow p_2\,,~~\label{3g-B3}
\eea
\bea
B_4&=&{\Big[S^{(1)_{p_3}}_g\,k_n\cdot f_{p_2}\cdot f_{p_1}\cdot Y_{p_1}\Big]\over k_n\cdot k_{p_1p_2}}\,{\cal A}_{\rm S}(1,\{2,\cdots,n-1\}\shuffle \{p_1,p_2\},n|\sigma_{n+3}\setminus p_3)+p_1\leftrightarrow p_2\nn
&=&\tau\,{k_n\cdot f_{p_3}\cdot f_{p_2}\cdot f_{p_1}\cdot Y_{p_1}\over k_n\cdot k_{p_1p_2}}\,{\cal A}^{(0)_{p_3}}_{\rm S}(1,\{2,\cdots,n-1\}\shuffle \{p_1,p_2,p_3\},n|\sigma_{n+3})+p_1\leftrightarrow p_2\nn
& &+\tau\,{k_n\cdot f_{p_2}\cdot f_{p_3}\cdot f_{p_1}\cdot Y_{p_1}\over k_n\cdot k_{p_1p_2}}\,{\cal A}^{(0)_{p_3}}_{\rm S}(1,\{2,\cdots,n-1\}\shuffle \{p_1,p_3,p_2\},n|\sigma_{n+3})+p_1\leftrightarrow p_2\nn
& &+\tau\,{k_n\cdot f_{p_2}\cdot f_{p_1}\cdot f_{p_3}\cdot Y_{p_3}\over k_n\cdot k_{p_1p_2}}\,{\cal A}^{(0)_{p_3}}_{\rm S}(1,\{2,\cdots,n-1\}\shuffle \{p_3,p_1,p_2\},n|\sigma_{n+3})+p_1\leftrightarrow p_2\,,~~\label{3g-B4}
\eea
\bea
B_5&=&\Big[S^{(1)_{p_3}}_g\,{1\over k_n\cdot k_{p_1p_2}}\Big]\,(k_n\cdot f_{p_1}\cdot Y_{p_1})\,{\cal A}_{\rm YS}(1,\{2,\cdots,n-1\}\shuffle p_1,n;p_2|\sigma_{n+3}\setminus p_3)+p_1\leftrightarrow p_2\nn
&=&-\tau\,{(k_n\cdot f_{p_1}\cdot Y_{p_1})\,(k_n\cdot f_{p_3}\cdot K_{p_3})\over \big(k_n\cdot k_{p_1p_2}\big)^2}\,\W{\cal A}^{(0)_{p_3}}_{\rm YS}(1,\{2,\cdots,n-1\}\shuffle p_1\shuffle p_3,n;p_2|\sigma_{n+3})+p_1\leftrightarrow p_2\,,~~\label{3g-B5}
\eea
\bea
B_6&=&\Big[S^{(1)_{p_3}}_g\,{1\over k_n\cdot k_{p_1p_2}}\Big]\,(k_n\cdot f_{p_2}\cdot f_{p_1}\cdot Y_{p_1})\,{\cal A}_{\rm S}(1,\{2,\cdots,n-1\}\shuffle \{p_1,p_2\},n|\sigma_{n+3}\setminus p_3)+p_1\leftrightarrow p_2\nn
&=&-\tau\,{(k_n\cdot f_{p_2}\cdot f_{p_1}\cdot Y_{p_1})\,(k_n\cdot f_{p_3}\cdot K_{p_3})\over \big(k_n\cdot k_{p_1p_2} \big)^2}\,{\cal A}^{(0)_{p_3}}_{\rm S}(1,\{2,\cdots,n-1\}\shuffle \{p_1,p_2\}\shuffle p_3,n|\sigma_{n+3}\setminus p_3)\nn
& &+p_1\leftrightarrow p_2\,.~~\label{3g-B6}
\eea
The first and second parts $B_1$ and $B_2$ in \eref{3g-B1} and \eref{3g-B2} are analogous to $B_1$ in \eref{B1}, come from the soft theorem. The computation of $B_3$ and $B_4$ in \eref{3g-B3} and \eref{3g-B4}
are parallel to that for obtaining $B_2$ in \eref{B2}. The calculation of $B_5$ and $B_6$ in \eref{3g-B5} and \eref{3g-B6} are parallel to the process for obtaining $B_3$ in \eref{B3}. The combinatory momentum $K_{p_3}$ is defined as the summation of momenta in $\{k_{p_1},k_{p_2}\}$
whose corresponding legs are at the l.h.s of $p_3$ in the color ordering. The notation $\W{\cal A}^{(0)_{p_3}}_{\rm YS}(1,\{2,\cdots,n-1\}\shuffle p_1\shuffle p_3,n;p_2|\sigma_{n+3})$ in the last line of \eref{3g-B5} means expanding ${\cal A}^{(0)_{p_3}}_{\rm YS}(1,\{2,\cdots,n-1\}\shuffle p_1\shuffle p_3,n;p_2|\sigma_{n+3})$ to pure BAS basis, and $K_{p_3}$ are defined for the corresponding color orderings in these BAS amplitudes. If one use other expressions for ${\cal A}^{(0)_{p_3}}_{\rm YS}(1,\{2,\cdots,n-1\}\shuffle p_1\shuffle p_3,n;p_2|\sigma_{n+3})$, \eref{3g-B5} does not hold.

Using $K_{p_3}=Y_{p_3}-X_{p_3}$, where for \eref{3g-B5} $X_{p_3}$ is defined for BAS amplitudes after expanding ${\cal A}^{(0)_{p_3}}_{\rm YS}(1,\{2,\cdots,n-1\}\shuffle p_1\shuffle p_3,n;p_2|\sigma_{n+3})$, one can separate $B_5$ and $B_6$ as
\bea
B_5=B_7+B_8\,,~~~~B_6=B_9+B_{10}\,,
\eea
where
\bea
B_7=\tau\,{(k_n\cdot f_{p_1}\cdot Y_{p_1})\,(k_n\cdot f_{p_3}\cdot Y_{p_3})\over \big(k_n\cdot k_{p_1p_2}\big)^2}\,{\cal A}^{(0)_{p_3}}_{\rm YS}(1,\{2,\cdots,n-1\}\shuffle p_1\shuffle p_3,n;p_2|\sigma_{n+3})+p_1\leftrightarrow p_2\,,~~\label{3g-B7}
\eea
\bea
B_8=-\tau\,{(k_n\cdot k_{p_3})\,(k_n\cdot f_{p_1}\cdot Y_{p_1})\over \big(k_n\cdot k_{p_1p_2}\big)^2}\,{\cal A}^{(0)_{p_3}}_{\rm YS}(1,\{2,\cdots,n-1\}\shuffle p_1,n;\{p_i\}_3\setminus p_1|\sigma_{n+3})+p_1\leftrightarrow p_2\,,~~\label{3g-B8}
\eea
\bea
B_9&=&\tau\,{(k_n\cdot f_{p_2}\cdot f_{p_1}\cdot Y_{p_1})\,(k_n\cdot f_{p_3}\cdot Y_{p_3})\over \big(k_n\cdot k_{p_1p_2}\big)^2}\,{\cal A}^{(0)_{p_3}}_{\rm S}(1,\{2,\cdots,n-1\}\shuffle \{p_1,p_2\}\shuffle p_3,n|\sigma_{n+3})\nn
& &+p_1\leftrightarrow p_2\,~~\label{3g-B9}
\eea
\bea
B_{10}=-\tau\,{(k_n\cdot k_{p_3})\,(k_n\cdot f_{p_2}\cdot f_{p_1}\cdot Y_{p_1})\over \big(k_n\cdot k_{p_1p_2}\big)^2}\,{\cal A}^{(0)_{p_3}}_{\rm YS}(1,\{2,\cdots,n-1\}\shuffle \{p_1,p_2\},n;p_3|\sigma_{n+3})+p_1\leftrightarrow p_2\,,~~\label{3g-B10}
\eea
The result $B_8$ in \eref{3g-B8} can be obtained as follows. Using the expansion of ${\cal A}_{\rm YS}(1,\{2,\cdots,n-1\}\shuffle p_1,n;\{p_i\}_3\setminus p_1|\sigma_{n+3})$, we have
\bea
& &{\cal A}^{(0)_{p_3}}_{\rm YS}(1,\{2,\cdots,n-1\}\shuffle p_1,n;\{p_i\}_3\setminus p_1|\sigma_{n+3})\nn
&=&{k_n\cdot f_{p_2}\cdot Y_{p_2}\over k_n\cdot k_{p_2}}\,{\cal A}^{(0)_{p_3}}_{\rm YS}(1,\{2,\cdots,n-1\}\shuffle p_1\shuffle p_2,n;p_3|\sigma_{n+3})\,,~~\label{B81}
\eea
since all other terms are at the $\tau^1$ order. Then we use the expansion of ${\cal A}_{\rm YS}(1,\{2,\cdots,n-1\}\shuffle p_1\shuffle p_2,n;p_3|\sigma_{n+3})$ to get
\bea
& &{\cal A}^{(0)_{p_3}}_{\rm YS}(1,\{2,\cdots,n-1\}\shuffle p_1\shuffle p_2,n;p_3|\sigma_{n+3})\nn
&=&{k_n\cdot f_{p_3}\cdot X_{p_3}\over k_n\cdot k_{p_3}}\,{\cal A}^{(0)_{p_3}}_{\rm S}(1,\{2,\cdots,n-1\}\shuffle p_1\shuffle p_2\shuffle p_3,n|\sigma_{n+3})\,.~~\label{B82}
\eea
Notice that for pure BAS amplitudes $Y_{p_3}$ is equivalent to $X_{p_3}$. Expanding ${\cal A}^{(0)_{p_3}}_{\rm YS}(1,\{2,\cdots,n-1\}\shuffle p_1\shuffle p_2,n;p_3|\sigma_{n+3})$ will not alter $Y_{p_2}$ in \eref{B81} at the leading order, since $k_{p_3}$
is accompanied with $\tau$ thus does not contribute. Substituting \eref{B82} into \eref{B81}, we find
\bea
& &{\cal A}^{(0)_{p_3}}_{\rm YS}(1,\{2,\cdots,n-1\}\shuffle p_1,n;\{p_i\}_3\setminus p_1|\sigma_{n+3})\nn
&=&{k_n\cdot f_{p_3}\cdot X_{p_3}\over k_n\cdot k_{p_3}}\,\W{\cal A}^{(0)_{p_3}}_{\rm YS}(1,\{2,\cdots,n-1\}\shuffle p_1\shuffle p_3,n;p_2|\sigma_{n+3})\,,~~\label{B83}
\eea
where the expansion of ${\cal A}_{\rm YS}(1,\{2,\cdots,n-1\}\shuffle p_1\shuffle p_3,n;p_2|\sigma_{n+3})$ has been used. The relation \eref{B83} yields the expression of $B_8$ in \eref{3g-B8}. The expression of $B_9$ in \eref{3g-B10} can be calculated in the analogous way.

Now one can recognize that $B_1$ in \eref{3g-B1} together with $B_8$ in \eref{3g-B8} gives the sub-leading term of
\bea
P_1={k_n\cdot f_{p_1}\cdot Y_{p_1}\over k_n\cdot k_{p_1p_2p_3}}\,{\cal A}_{\rm YS}(1,\{2,\cdots,n-1\}\shuffle p_1,n;\{p_i\}_3\setminus p_1|\sigma_{n+3})+p_1\leftrightarrow p_2\,,~~\label{3g-P1}
\eea
while $B_2$ in \eref{3g-B2} and $B_{10}$ in \eref{3g-B10} gives the sub-leading term of
\bea
P_2={k_n\cdot f_{p_2}\cdot f_{p_1}\cdot Y_{p_1}\over k_n\cdot k_{p_1p_2p_3}}\,{\cal A}_{\rm YS}(1,\{2,\cdots,n-1\}\shuffle \{p_1,p_2\},n;p_3|\sigma_{n+3})+p_1\leftrightarrow p_2\,.~~\label{3g-P2}
\eea
Similar as in the $2$-gluon case, $k_n\cdot k_{p_1p_2}$ in the denominator is turned to $k_{n}\cdot k_{p_1p_2p_3}$, to keep the manifest permutation symmetry among external gluons $p_1$, $p_2$, $p_3$. The block $B_3$ in \eref{3g-B3} can be interpreted as the leading term of
\bea
P_3&=&{k_n\cdot f_{p_3}\cdot f_{p_1}\cdot Y_{p_1}\over k_n\cdot k_{p_1p_2p_3}}\,{\cal A}_{\rm YS}(1,\{2,\cdots,n-1\}\shuffle \{p_1,p_3\},n;p_2|\sigma_{n+3})+p_1\leftrightarrow p_2\nn
& &+{k_n\cdot f_{p_1}\cdot f_{p_3}\cdot Y_{p_3}\over k_n\cdot k_{p_1p_2p_3}}\,{\cal A}_{\rm YS}(1,\{2,\cdots,n-1\}\shuffle \{p_3,p_1\},n;p_2|\sigma_{n+3})+p_1\leftrightarrow p_2\,,~~\label{3g-B3}
\eea
$B_4$ in \eref{3g-B4} can be interpreted as the leading term of
\bea
P_4&=&{k_n\cdot f_{p_3}\cdot f_{p_2}\cdot f_{p_1}\cdot Y_{p_1}\over k_n\cdot k_{p_1p_2p_3}}\,{\cal A}_{\rm S}(1,\{2,\cdots,n-1\}\shuffle \{p_1,p_2,p_3\},n|\sigma_{n+3})+p_1\leftrightarrow p_2\nn
& &+{k_n\cdot f_{p_2}\cdot f_{p_3}\cdot f_{p_1}\cdot Y_{p_1}\over k_n\cdot k_{p_1p_2p_3}}\,{\cal A}_{\rm S}(1,\{2,\cdots,n-1\}\shuffle \{p_1,p_3,p_2\},n|\sigma_{n+3})+p_1\leftrightarrow p_2\nn
& &+{k_n\cdot f_{p_2}\cdot f_{p_1}\cdot f_{p_3}\cdot Y_{p_3}\over k_n\cdot k_{p_1p_2p_3}}\,{\cal A}_{\rm S}(1,\{2,\cdots,n-1\}\shuffle \{p_3,p_1,p_2\},n|\sigma_{n+3})+p_1\leftrightarrow p_2\,.~~\label{3g-P4}
\eea
Finally, combining $B_7$ in \eref{3g-B7} and $B_9$ in \eref{3g-B9} together gives the leading contribution of
\bea
P_5={k_n\cdot f_{p_3}\cdot Y_{p_3}\over k_n\cdot k_{p_1p_2p_3}}\,{\cal A}_{\rm YS}(1,\{2,\cdots,n-1\}\shuffle p_3,n;\{p_i\}_2|\sigma_{n+3})\,.~~\label{3g-P5}
\eea
Thus, we arrive at
\bea
& &{\cal A}_{\rm YS}(1,\cdots,n;\{p_i\}_3|\sigma_{n+3})\nn
&=&P_1+P_2+P_3+P_4+P_5\nn
&=&\sum_{\a_1\in\{p_i\}_3}\,{k_n\cdot f_{\a_1}\cdot Y_{\a_1}\over k_n\cdot k_{p_1p_2p_3}}\,{\cal A}_{\rm YS}(1,\{2,\cdots,n-1\}\shuffle \a_1,n;\{p_i\}_3\setminus\a_1|\sigma_{n+3})\nn
& &+\sum_{\b_1,\b_2\in\{p_i\}_3}\,{k_n\cdot f_{\b_2}\cdot f_{\b_1}\cdot Y_{\b_1}\over k_n\cdot k_{p_1p_2p_3}}\,{\cal A}_{\rm YS}(1,\{2,\cdots,n-1\}\shuffle \{\b_1,\b_2\},n;\{p_i\}_3\setminus\{\b_1,\b_2\}|\sigma_{n+3})\nn
& &+\sum_{\gamma_j\in\{p_i\}_3}\,{k_n\cdot f_{\gamma_3}\cdot f_{\gamma_2}\cdot f_{\gamma_1}\cdot Y_{\b_1}\over k_n\cdot k_{p_1p_2p_3}}\,{\cal A}_{\rm S}(1,\{2,\cdots,n-1\}\shuffle \{\gamma_1,\gamma_2,\gamma_3\},n|\sigma_{n+3})\,,~~\label{expan-3g-gi}
\eea
where $j\in\{1,2,3\}$ in the final line.

\subsection{General case}
\label{subsec-general}

Comparing the constructions in subsections.\ref{subsec-2gluon} and \ref{subsec-3gluon}, we see the processes bear strong similarity. This similarity implies the recursive pattern which yields the general expansion for single trace YMS amplitudes with arbitrary number of external gluons. Such general expansion is the goal of this subsection.

In general, the single-trace YMS amplitude ${\cal A}_{\rm YS}(1,\cdots,n;\{p_i\}_m|\sigma_{n+m})$ with $m$ external gluons can be expanded as
in \eref{expan-YMS-gi}. When $\pmb{\a}=\{p_i\}_m$, the YMS amplitudes at the second line of \eref{expan-YMS-gi} are reduced to pure BAS ones. As mentioned previously, in this note we restrict ourselves to the special choice of the reference momentum $k_r$ in \eref{expan-YMS-gi}, which is $k_r=k_n$.

Obviously, expansions in \eref{expan-1g-gi}, \eref{expan-2g-gi}, \eref{expan-3g-gi} satisfy the general formula in \eref{expan-YMS-gi} with $k_r=k_n$. To achieve the general formula, let us prove that suppose \eref{expan-YMS-gi} is satisfied for a particular $m$, then it is also satisfied for $m+1$. Consider $k_{p_{m+1}}\to\tau k_{p_{m+1}}$, and expand ${\cal A}_{\rm YS}(1,\cdots,n;\{p_i\}_{m+1}|\sigma_{n+m+1})$ in $\tau$ to get
\bea
{\cal A}^{(1)_{p_{m+1}}}_{\rm YS}(1,\cdots,n;\{p_i\}_{m+1}|\sigma_{n+m+1})
&=&S^{(1)_{p_{m+1}}}_g\,{\cal A}_{\rm YS}(1,\cdots,n;\{p_i\}_m|\sigma_{n+m+1}\setminus p_{m+1})\nn
&=&B_1+B_2+B_3\,,~~\label{3B}
\eea
where
\bea
B_1&=&\sum_{\vec{\pmb{\a}}}\,{k_n\cdot F_{\vec{\pmb{\a}}}\cdot Y_{\vec{\pmb{\a}}}\over k_n\cdot k_{p_1\cdots p_m}}\nn
& &\Big[S^{(1)_{p_{m+1}}}_g\,{\cal A}_{\rm YS}(1,\{2,\cdots,n-1\}\shuffle\vec{\pmb{\a}},n;\{p_i\}_m\setminus\pmb{\a}|\sigma_{n+m+1}\setminus p_{m+1})\Big]\nn
&=&\sum_{\vec{\pmb{\a}}}\,{k_n\cdot F_{\vec{\pmb{\a}}}\cdot Y_{\vec{\pmb{\a}}}\over k_n\cdot k_{p_1\cdots p_m}}\nn
& &{\cal A}^{(1)_{p_{m+1}}}_{\rm YS}(1,\{2,\cdots,n-1\}\shuffle\vec{\pmb{\a}},n;\{p_i\}_{m+1}\setminus\pmb{\a}|\sigma_{n+m+1})\,,~~\label{mg-B1}
\eea
\bea
B_2
&=&\sum_{\vec{\pmb{\a}}}\,{\Big[S^{(1)_{p_{m+1}}}_g\,k_n\cdot F_{\vec{\pmb{\a}}}\cdot Y_{\vec{\pmb{\a}}}\Big]\over k_n\cdot k_{p_1\cdots p_m}}\nn
& &{\cal A}_{\rm YS}(1,\{2,\cdots,n-1\}\shuffle\vec{\pmb{\a}},n;\{p_i\}_m\setminus\pmb{\a}|\sigma_{n+m+1}\setminus p_{m+1})\nn
&=&\sum_{\vec{\pmb{\a}}'}\,\tau\,{k_n\cdot F_{\vec{\pmb{\a}}'}\cdot Y_{\vec{\pmb{\a}}'}\over k_n\cdot k_{p_1\cdots p_m}}\nn
& &{\cal A}^{(0)_{p_{m+1}}}_{\rm YS}(1,\{2,\cdots,n-1\}\shuffle\vec{\pmb{\a}}',n;\{p_i\}_{m+1}\setminus\pmb{\a'}|\sigma_{n+m+1})\,,~~\label{mg-B2}
\eea
with $\vec{\pmb{\a}}'=\vec{\pmb{\a}}\shuffle p_{m+1}$, and
\bea
B_3
&=&\sum_{\vec{\pmb{\a}}}\,\Big[S^{(1)_{p_{m+1}}}_g\,{1\over k_n\cdot k_{p_1\cdots p_m}}\Big]\,(k_n\cdot F_{\vec{\pmb{\a}}}\cdot Y_{\vec{\pmb{\a}}})\nn
& &{\cal A}_{\rm YS}(1,\{2,\cdots,n-1\}\shuffle\vec{\pmb{\a}},n;\{p_i\}_m\setminus\pmb{\a}|\sigma_{n+m+1}\setminus p_{m+1})\nn
&=&-\sum_{\vec{\pmb{\a}}}\,\tau\,{(k_n\cdot F_{\vec{\pmb{\a}}}\cdot Y_{\vec{\pmb{\a}}})\,(k_n\cdot f_{p_{m+1}}\cdot K_{p_{m+1}})\over \big(k_n\cdot k_{p_1\cdots p_m}\big)^2}\nn
& &\W{\cal A}^{(0)_{p_{m+1}}}_{\rm YS}(1,\{2,\cdots,n-1\}\shuffle\vec{\pmb{\a}}\shuffle p_{m+1},n;\{p_i\}_m\setminus\pmb{\a}|\sigma_{n+m+1})\,.~~\label{mg-B3}
\eea
The calculation are paralleled to those in subsections \ref{subsec-2gluon} and \ref{subsec-3gluon}. Again, the notation $\W{\cal A}^{(0)_{p_{m+1}}}_{\rm YS}(1,\{2,\cdots,n-1\}\shuffle\vec{\pmb{\a}}\shuffle p_{m+1},n;\{p_i\}_m\setminus\pmb{\a}|\sigma_{n+m+1})$ is understood as expanding ${\cal A}^{(0)_{p_{m+1}}}_{\rm YS}(1,\{2,\cdots,n-1\}\shuffle\vec{\pmb{\a}}\shuffle p_{m+1},n;\{p_i\}_m\setminus\pmb{\a}|\sigma_{n+m+1})$ to BAS basis. To obtain $B_2$ in the second line of \eref{mg-B2}, we used the property
\bea
\vec{\pmb{a}}\shuffle\vec{\pmb{b}}\shuffle\vec{\pmb{c}}=\vec{\pmb{a}}\shuffle\{\vec{\pmb{b}}\shuffle\vec{\pmb{c}}\}\,,
\eea
where $\vec{\pmb{a}}$, $\vec{\pmb{b}}$ and $\vec{\pmb{c}}$ are three ordered sets. In \eref{mg-B3}, $K_{p_{m+1}}$ stands for the summation of momenta carried by external gluons $p_i$ at the l.h.s of $p_{m+1}$ in the color ordering. We emphasize that the external leg $p_{m+1}$ is not included in the set $\pmb{\a}$ appear
in $B_1$, $B_2$ and $B_3$.

Using $K_{p_{m+1}}=Y_{p_{m+1}}-X_{p_{m+1}}$, one can split $B_3$ as
\bea
B_3=B_4+B_5\,,
\eea
where
\bea
B_4&=&\sum_{\vec{\pmb{\a}}}\,\tau\,{(k_n\cdot F_{\vec{\pmb{\a}}}\cdot Y_{\vec{\pmb{\a}}})\,(k_n\cdot f_{p_{m+1}}\cdot Y_{p_{m+1}})\over \big(k_n\cdot k_{p_1\cdots p_m}\big)^2}\nn
& &{\cal A}^{(0)_{p_{m+1}}}_{\rm YS}(1,\{2,\cdots,n-1\}\shuffle\vec{\pmb{\a}}\shuffle p_{m+1},n;\{p_i\}_m\setminus\pmb{\a}|\sigma_{n+m+1})\nn
&=&\tau\,{k_n\cdot f_{p_{m+1}}\cdot Y_{p_{m+1}}\over k_n\cdot k_{p_1\cdots p_m}}\,{\cal A}^{(0)_{p_{m+1}}}_{\rm YS}(1,\{2,\cdots,n-1\}\shuffle p_{m+1},n;\{p_i\}_m|\sigma_{n+m+1})\,,~~\label{mg-B4}
\eea
\bea
B_5&=&-\sum_{\vec{\pmb{\a}}}\,\tau\,{(k_n\cdot F_{\vec{\pmb{\a}}}\cdot Y_{\vec{\pmb{\a}}})\,(k_n\cdot f_{p_{m+1}}\cdot X_{p_{m+1}})\over \big(k_n\cdot k_{p_1\cdots p_m}\big)^2}\nn
& &\W{\cal A}^{(0)_{p_{m+1}}}_{\rm YS}(1,\{2,\cdots,n-1\}\shuffle\vec{\pmb{\a}}\shuffle p_{m+1},n;\{p_i\}_m\setminus\pmb{\a}|\sigma_{n+m+1})\nn
&=&-\sum_{\vec{\pmb{\a}}}\,\tau\,{(k_n\cdot k_{p_{m+1}})\,(k_n\cdot F_{\vec{\pmb{\a}}}\cdot Y_{\vec{\pmb{\a}}})\over \big(k_n\cdot k_{p_1\cdots p_m}\big)^2}\nn
& &{\cal A}^{(0)_{p_{m+1}}}_{\rm YS}(1,\{2,\cdots,n-1\}\shuffle\vec{\pmb{\a}},n;\{p_i\}_{m+1}\setminus\pmb{\a}|\sigma_{n+m+1})\,.~~\label{mg-B5}
\eea
The computations of $B_4$ and $B_5$ are analogous to those for obtaining $B_7$, $B_8$, $B_9$ and $B_{10}$ in subsection. \ref{subsec-3gluon}. It is worth to give a brief discussion for how to generalize the manipulation from \eref{B81} to \eref{B83} to the current general case. We first observe that
\bea
& &{\cal A}^{(0)_{p_{m+1}}}_{\rm YS}(1,\{2,\cdots,n-1\}\shuffle\vec{\pmb{\a}},n;\{p_i\}_{m+1}\setminus\pmb{\a}|\sigma_{n+m+1})\nn
&=&\sum_{\vec{\pmb{\b}}}\,{k_n\cdot F_{\vec{\pmb{\b}}}\cdot Y_{\vec{\pmb{\b}}}\over k_n\cdot k_{\{p_i\}_{m}\setminus\pmb{\a}}}\,
{\cal A}^{(0)_{p_{m+1}}}_{\rm YS}(1,\{2,\cdots,n-1\}\shuffle\vec{\pmb{\a}}\shuffle\vec{\pmb{\b}},n;\{p_i\}_{m+1}\setminus\{\pmb{\a}\cup\pmb{\b}\}|\sigma_{n+m+1})\,.
~~\label{toB5-1}
\eea
Here $\pmb{\b}$ is a subset of $\{p_i\}_m\setminus\pmb{\a}$ which does not include $p_{m+1}$, since otherwise $k_{p_{m+1}}$ will occur in $F_{\vec{\pmb{\b}}}$ and contributes $\tau$ under the re-scaling $k_{p_{m+1}}\to\tau k_{p_{m+1}}$. The combinatory momentum $k_{\{p_i\}_m\setminus\pmb{\a}}$ is the summation of momenta for external legs in the set $\{p_i\}_m\setminus\pmb{\a}$. Applying the above procedure iteratively, one can expand the leading terms of YMS amplitudes in the second line of \eref{toB5-1} further, until there is only one remaining gluon $p_{m+1}$. In other words, ${\cal A}^{(0)_{p_{m+1}}}_{\rm YS}(1,\{2,\cdots,n-1\}\shuffle\vec{\pmb{\a}},n;\{\{p_i\}_{m+1}\setminus\pmb{\a}\}|\sigma_{n+m+1})$ can be expanded to
${\cal A}^{(0)_{p_{m+1}}}_{\rm YS}(1,\{2,\cdots,n-1\}\shuffle p_1\shuffle\cdots\shuffle p_m,n;p_{m+1}|\sigma_{n+m+1})$. Then, we expand them further by employing
\bea
& &{\cal A}^{(0)_{p_{m+1}}}_{\rm YS}(1,\{2,\cdots,n-1\}\shuffle p_1\shuffle\cdots\shuffle p_m,n;p_{m+1}|\sigma_{n+m+1})\nn
&=&{k_n\cdot f_{p_{m+1}}\cdot X_{p_{m+1}}\over k_n\cdot k_{p_{m+1}}}\,{\cal A}^{(0)_{p_{m+1}}}_{\rm S}(1,\{2,\cdots,n-1\}\shuffle p_1\shuffle\cdots\shuffle p_m\shuffle p_{m+1},n|\sigma_{n+m+1})\,,~~\label{toB5-2}
\eea
and reorganize the full expansion as
\bea
& &{\cal A}^{(0)_{p_{m+1}}}_{\rm YS}(1,\{2,\cdots,n-1\}\shuffle\vec{\pmb{\a}},n;\{p_i\}_{m+1}\setminus\pmb{\a}|\sigma_{n+m+1})\nn
&=&{k_n\cdot f_{p_{m+1}}\cdot X_{p_{m+1}}\over k_n\cdot k_{p_{m+1}}}\,\W{\cal A}^{(0)_{p_{m+1}}}_{\rm YS}(1,\{2,\cdots,n-1\}\shuffle\vec{\pmb{\a}}\shuffle p_{m+1},n;\{p_i\}_m\setminus\pmb{\a}|\sigma_{n+m+1})\,,~~\label{toB5-3}
\eea
based on the observation that $k_{p_{m+1}}$ which is accompanied with $\tau$ does not contribute to any $Y$ at the leading order.
The relation \eref{toB5-3} is the generalization of the previous result \eref{B83}.
Using \eref{toB5-3}, we arrive at the expression of $B_5$ in \eref{mg-B5}.

Now we observe that $B_1$ in \eref{mg-B1} together with $B_5$ in \eref{mg-B5} give the sub-leading term of
\bea
P_1=\sum_{\vec{\pmb{\a}}}\,{k_n\cdot F_{\vec{\pmb{\a}}}\cdot Y_{\vec{\pmb{\a}}}\over k_n\cdot k_{p_1\cdots p_{m+1}}}\,
{\cal A}_{\rm YS}(1,\{2,\cdots,n-1\}\shuffle\vec{\pmb{\a}},n;\{p_i\}_{m+1}\setminus\pmb{\a}|\sigma_{n+m+1})\,,~~\label{mg-P1}
\eea
$B_2$ in \eref{mg-B2} can be interpreted as the leading term of
\bea
P_2=\sum_{\vec{\pmb{\a}}'}\,{k_n\cdot F_{\vec{\pmb{\a}}'}\cdot Y_{\vec{\pmb{\a}}'}\over k_n\cdot k_{p_1\cdots p_{m+1}}}\,
{\cal A}_{\rm YS}(1,\{2,\cdots,n-1\}\shuffle\vec{\pmb{\a}}',n;\{p_i\}_{m+1}\setminus\pmb{\a'}|\sigma_{n+m+1})\,,~~\label{mg-P2}
\eea
where $\pmb{\a}'$ includes the external leg $p_{m+1}$ and $\pmb{\a}'\setminus p_{m+1}\neq\emptyset$, and $B_4$ in \eref{mg-B4} corresponds to the leading terms of
\bea
P_3={k_n\cdot f_{p_{m+1}}\cdot Y_{p_{m+1}}\over k_n\cdot k_{p_1\cdots p_{m+1}}}\,{\cal A}_{\rm YS}(1,\{2,\cdots,n-1\}\shuffle p_{m+1},n;\{p_i\}_m|\sigma_{n+m+1})\,.~~\label{mg-P3}
\eea
Combining $P_1$, $P_2$ and $P_3$ together, we see that the general formula \eref{expan-YMS-gi} is correct for single-trace YMS amplitude with $(m+1)$
external gluons, thus the proof is completed.

\section{Summery}
\label{sec-conclusion}

In this note, we reconstructed the expansion of single-trace YMS amplitudes proposed by Clifford Cheung and James Mangan in \cite{Cheung:2021zvb}, via the bottom up method based on soft theorems, with out the aid of a Lagrangian or equations of motion. The whole process is on-shell, without using any off-shell ansatz. The new recursive method inserts gluons into the original amplitude in the manifestly gauge invariant manner, leads to the resulting expansion which manifests the gauge invariance for all polarizations carried by external gluons, and the permutation symmetry among external gluons, with the cost of sacrificing the manifest locality. One can use such expansion iteratively to expand the single-trace YMS amplitudes to pure BAS ones without breaking the manifest gauge invariance for any polarization, but a verity of spurious poles will be created.

Through the double copy structure, our result can be extended to the expansion of single-trace Einstein-Yang-Mills amplitudes directly,
\bea
& &{\cal A}_{\rm EY}(1,\cdots,n;\{p_i\}_m)\nn
&=&\sum_{\vec{\pmb{\a}}}\,{k_n\cdot F_{\vec{\pmb{\a}}}\cdot Y_{\vec{\pmb{\a}}}\over k_n\cdot k_{p_1\cdots p_m}}\,{\cal A}_{\rm EY}(1,\{2,\cdots,n-1\}\shuffle\vec{\pmb{\a}},n;\{p_i\}_m\setminus{\pmb{\a}})\,,~~\label{expan-EY-gi}
\eea
where $1,\cdots,n$ are color ordered gluons, and elements in $\{p_i\}_m$ are gravitons without any ordering. It is straightforward to verify
the expansion \eref{expan-EY-gi} is equivalent to that found by Clifford Cheung and James Mangan in \cite{Cheung:2021zvb}. After some modifications of notations, the expansion in \cite{Cheung:2021zvb} with the choice of reference momentum $k_r=k_n$ is given as
\bea
& &{\cal A}_{\rm YS}(1,\cdots,n;\{p_i\}_m)\nn
&=&\sum_{\vec{\pmb{\a}}}\,\sum_{i=1}^{n-1}\,{k_n\cdot F_{\vec{\pmb{\a}}}\cdot k_i\over k_n\cdot k_{1\cdots n}}\,{\cal A}_{\rm YS}(1,\cdots,i,\{i+1,\cdots,n-1\}\shuffle\vec{\pmb{\a}},n;\{p_i\}_m\setminus{\pmb{\a}})\,,~~\label{expan-EY-gi-cheung}
\eea
up to an overall sign.
Here $k_i$ contributes when $i$ is at the l.h.s of the first element in $\vec{\pmb{\a}}$ in the ordering, therefore the definition of $Y_{\vec{\pmb{\a}}}$ is satisfied. On the other hand, the momentum conservation gives $k_{1\cdots n}=-k_{p_1\cdots p_m}$. Thus the equivalence between \eref{expan-EY-gi} and \eref{expan-EY-gi-cheung} is proved.

The sub-leading soft theorem for external gluon plays the crucial role in our recursive method. One may ask the reason for choosing sub-leading soft theorem rather than the leading order one. The answer is, the leading order soft theorem can not detect all terms in the expansion. Suppose we re-scale $k_{p_j}$ as $k_{p_j}\to\tau k_{p_j}$, as can be seen in the general expanded formula \eref{expan-YMS-gi}, various terms include $f_{p_j}^{\mu\nu}$ in corresponding coefficients therefore are accompanied with $\tau$, these terms have no contribution to the leading order.
Another question is whether the sub-leading soft theorem is sufficient to detect all terms in the complete expansion? For example, if the coefficient of one term behaviors as $\tau^2$, then this coefficient can not be detected at both leading and sub-leading orders. This question was answered in \cite{Zhou:2022orv}, by employing the universality of soft factor for external scalars, as well as the counting of mass dimensions. The argument in \cite{Zhou:2022orv} at least shows that the old expansion \eref{expan-YMS-recur} can be fully detected by the sub-leading soft theorem for external gluon. Since any correct expansion must be equivalent to \eref{expan-YMS-recur}, we claim that the new expansion, which manifests gauge invariance for all polarizations, can also be detected completely.

A related interesting future direction is seeking the expanded formula for pure YM amplitudes which manifests the gauge invariance for each polarization. Such expansion leads to the manifestly gauge invariant BCJ numerators. We expect the recursive method in this note can be applied to solve this challenge.

\section*{Acknowledgments}

The authors would thank Prof.Yijian Du for stimulating discussions. We also thank Dr.Linghui Hou who told us the expansion found by Clifford Cheung and James Mangan.


\end{document}